\documentclass[aps,pra,reprint,amsmath,amssymb,superscriptaddress,onecolumn,longbibliography,nofootinbib,notitlepage]{revtex4-2}
\usepackage{mathtools}
\usepackage{siunitx}
\DeclareSIUnit{\sccm}{sccm}
\DeclareSIUnit{\torr}{Torr}
\usepackage[hidelinks]{hyperref}
\usepackage{svg}
\usepackage{caption}
\usepackage{float}
\usepackage{subcaption}
\usepackage[braket, qm]{qcircuit}
\usepackage{bbm}

\usepackage{csquotes}
\usepackage{multirow}
\usepackage{array}
\usepackage{pbox}

\newcommand{\im}{\mathrm{i}}
\newcommand{\pder}[2]{\frac{\partial#1}{\partial#2}}
\newcommand{\unitforN}{mode}

\usepackage{xcolor}

\makeatletter

\newcommand{\authormarks}[1]{%
  \@AF@join{\textsuperscript{#1}}%
}

\renewcommand{\doauthor}[3]{%
  \ignorespaces#1\unskip\@listcomma
  \begingroup
    #3%
  \@if@empty{#2}
    {\endgroup{}{}}
    {\endgroup{\comma@space}{}#2}%
  \space\@listand
}

\newcommand{\authornote}[2]{%
  \g@addto@macro\@FMN@list{\@TBN@opr{#1}{#2}}%
}

\makeatother

\begin{document}

\title{Lithography-free in situ photonic inverse design}

\author{Yiqi~Zhao}
\authormarks{*,\textsection}
\affiliation{School of Applied and Engineering Physics, Cornell University, Ithaca, NY 14853, USA}

\author{Martin~M.~Stein}
\authormarks{*,\textdagger,\textsection}
\affiliation{School of Applied and Engineering Physics, Cornell University, Ithaca, NY 14853, USA}
\affiliation{Physics \& Informatics Laboratories, NTT Research, Inc., Sunnyvale, CA 94085, USA}

\author{Fan~O.~Wu}
\affiliation{School of Applied and Engineering Physics, Cornell University, Ithaca, NY 14853, USA}

\author{Ryotatsu~Yanagimoto}
\affiliation{School of Applied and Engineering Physics, Cornell University, Ithaca, NY 14853, USA}
\affiliation{Physics \& Informatics Laboratories, NTT Research, Inc., Sunnyvale, CA 94085, USA}

\author{Benjamin~A.~Ash}
\affiliation{School of Applied and Engineering Physics, Cornell University, Ithaca, NY 14853, USA}

\author{Mandar~M.~Sohoni}
\affiliation{School of Applied and Engineering Physics, Cornell University, Ithaca, NY 14853, USA}

\author{Logan~G.~Wright}
\authormarks{\textdaggerdbl}
\affiliation{School of Applied and Engineering Physics, Cornell University, Ithaca, NY 14853, USA}
\affiliation{Physics \& Informatics Laboratories, NTT Research, Inc., Sunnyvale, CA 94085, USA}

\author{Tatsuhiro~Onodera}
\affiliation{School of Applied and Engineering Physics, Cornell University, Ithaca, NY 14853, USA}
\affiliation{Physics \& Informatics Laboratories, NTT Research, Inc., Sunnyvale, CA 94085, USA}

\author{Peter~L.~McMahon}
\authormarks{\textsection}
\affiliation{School of Applied and Engineering Physics, Cornell University, Ithaca, NY 14853, USA}
\affiliation{Kavli Institute at Cornell for Nanoscale Science, Cornell University, Ithaca, NY 14853, USA}

\authornote{1}{%
  Y.Z. and M.M. Stein contributed equally to this work.%
}

\authornote{2}{%
  Present addresses: Department of Applied Physics, Yale University,
  New Haven, CT 06520, USA; Physics \& Informatics Laboratories,
  NTT Research, Inc., Sunnyvale, CA 94085, USA.%
}

\authornote{3}{%
  Present address: Department of Applied Physics, Yale University,
  New Haven, CT 06520, USA.%
}

\authornote{4}{%
  Correspondence to:
  \href{mailto:yz2893@cornell.edu}{yz2893@cornell.edu};
  \href{mailto:ms3452@cornell.edu}{ms3452@cornell.edu};
  \href{mailto:pmcmahon@cornell.edu}{pmcmahon@cornell.edu}.%
}

\begin{abstract}
Inverse design has emerged as a powerful engineering approach in nanophotonics, leading to devices with unintuitive designs that extend beyond human intuition. However, inverse-designed devices have been limited by costly and time-consuming fabrication iterations, simulation--reality gaps, and the fixed nature of etched devices. Separately, programmable photonics has emerged as an approach to modify the function of photonic devices after fabrication; recent free-form programmable-photonic devices have been scaled to have many degrees of freedom, in principle providing platforms well-suited for programmably realizing inverse design of conventional linear-photonic functions. Here we demonstrate in situ training of a single free-form programmable-photonic device to achieve a variety of functions that each would have been implemented in separate devices in the conventional inverse-design paradigm. Our platform is a thin-film lithium niobate slab waveguide with full programmability over the refractive index of the two-dimensional slab, with a tuning range of up to $1.7\times10^{-3}$ for each of the $10^4$ effective pixels. We demonstrate $4\times4$ Gaussian-mode permutations, $4\times4$ Hermite--Gaussian mode demultiplexing, and $8\times8$ Gaussian-mode permutations, with maximum crosstalk of \SI{-27}{\decibel}, \SI{-13}{\decibel}, and \SI{-13}{\decibel}, as well as $4\times4$ matrix--vector multiplication. We further demonstrate engineering of spectral responses, showcasing both wavelength-independent operation throughout the \SIrange{1500}{1600}{\nano\metre} window and wavelength demultiplexing between \SI{1310}{\nano\metre} and \SI{1550}{\nano\metre}, with maximum crosstalk of \SI{-8.4}{\decibel}. Since the device is reconfigured for each function after fabrication, the investment in device fabrication, characterization, and calibration can be amortized across the various photonic functions the device is ultimately used for. The combination of programmable multimode wave propagation and inverse design provides an alternative paradigm for integrated photonics where spatial and spectral degrees of freedom can be manipulated simultaneously; where manufacturing and environmental imperfections can be mitigated dynamically; and where a single fabricated device can realize and switch between many different analog functions.

\end{abstract}

\maketitle

\begin{figure}[!htbp]
    \centering
    \includegraphics[width=1\linewidth]{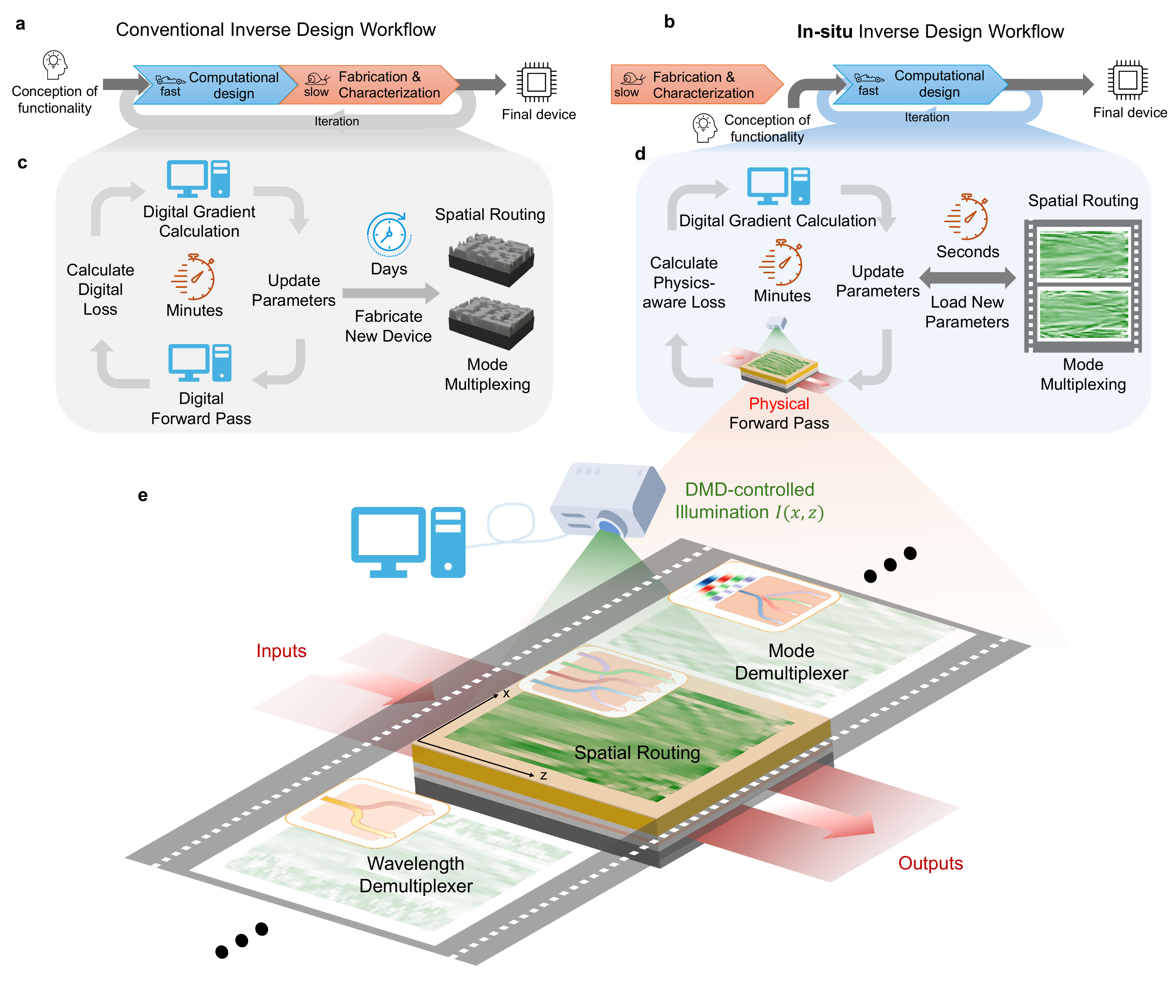}
    \caption{\textbf{Conventional vs. in situ photonic inverse design process, enabled by a 2D-programmable slab waveguide.} \textbf{a}, Workflow for conventional photonic inverse design. Designs are created by optimizing a digital simulation of the device, then fabricated and characterized to measure the performance. To obtain devices with different functions or to iterate on designs, one needs to repeat the complete process including slow fabrication steps. \textbf{b}, Workflow for in situ photonic inverse design. In contrast to the conventional approach, we started by fabricating and carefully characterizing a two-dimensionally programmable slab waveguide. Subsequently, we optimized the programmable parameters in situ, in an optical setup, inverse designing a device and simultaneously characterizing its performance. To achieve multiple functions, we can run different inverse design processes without fabricating additional devices. \textbf{c}, Inverse design loop and fabrication process for conventional inverse design. The inverse design workflow is performed digitally, with optimization times ranging from minutes to days. The optimized layouts are then fabricated, which typically requires at least hours to days. Because each design is fixed after fabrication, multiple devices need to be fabricated to realize different functions. \textbf{d}, Inverse design loop and fabrication process for in situ inverse design. We use a slab waveguide whose function is defined by the illumination pattern projected onto it. We use a hybrid physical-digital optimization algorithm, incorporating physical propagation of light through the device, using experimentally grounded feedback to find optimal illumination patterns. Optimization usually takes several minutes, after which the device is fully functional. We can implement various functions on a single device by switching between illumination patterns, enabling real-time reconfiguration without additional fabrication in seconds. \textbf{e}, Schematic of our 2D-programmable slab waveguide. We can change illumination patterns to display devices with different functions, with examples in the figure showing patterns for mode demultiplexing, spatial routing, and wavelength demultiplexing. DMD: digital micromirror device.}
    \label{fig:1}
\end{figure}

\section{Introduction}
\label{sec:intro}

Photonic inverse design is a powerful technique to discover non-intuitive designs to achieve specific photonic functions~\cite{moleskyInverseDesignNanophotonics2018, jensenTopologyOptimizationNanophotonics2011,gertlerManyPhotonicDesign2025}. Inverse design is performed by having a computer perform optimization over a design space and attempt to minimize a cost function that specifies the behavior that the inverse-designed device should have. There have been many successful demonstrations of inverse design, including for spatial routing~\cite{luInversedesignedUltracompactStarcrossings2017, xu5x5Mm2Compact2017, shangInverseDesignedLithiumNiobate2023}, mode multiplexing~\cite{changUltracompactModeMultiplexer2018, xieHighlyCompactEfficient2020, sunEdgeguidedInverseDesign2025, wuCompactScalableMode2025}, mode conversion~\cite{douUltracompactUltrabroadbandArbitraryorder2023, jiaUltracompactDualpolarizationSilicon2019}, wavelength multiplexing~\cite{piggottInverseDesignDemonstration2015, suInverseDesignDemonstration2018, cheungInversedesignedCWDMDemultiplexer2024, masonHighPerformanceWavelengthDivision2025}, and polarization control~\cite{changInverseDesignSinglestepetched2020}.

Despite these advances, most inverse-designed devices have been limited to small-scale demonstrations, with large-scale photonic integrated circuits still typically built from human-designed components. A central challenge in inverse design is that simulations of device behavior using a digital model typically do not accurately match the experimentally measured results of a device after it has been fabricated. This simulation--reality gap between digital models and the fabricated devices generally results in worse performance of inverse-designed devices than what simulations predict~\cite{nikkhahInversedesignedLowindexcontrastStructures2024}. Simulation times of hours~\cite{sunEdgeguidedInverseDesign2025,kangLargescalePhotonicInverse2024} and fabrication times of hours to weeks~\cite{bogaertsProgrammablePhotonicCircuits2020} are both costly and limit the speed of iteration for improving inverse-designed devices.

One way of addressing the simulation--reality gap is to introduce programmability into the physical device, allowing the parameters that control the device's function to be tuned post-fabrication and optimized using knowledge of the actual behavior in experiment. Programmable integrated photonic platforms typically allow tunability of degrees of freedom within human-designed devices, including Mach--Zehnder interferometers (MZIs)~\cite{bogaertsProgrammablePhotonicCircuits2020} and microring resonators (MRRs)~\cite{yiMultifunctionalPhotonicProcessors2021}. However, free-form programmable photonic platforms---such as with photorefractive crystals~\cite{Psaltis1990holography, Brady1991holographic}---have recently been revisited with phase-change materials (PCMs)~\cite{delaney-muskens2021nonvolatile, wu-li2024freeform, wuReconfigurableInversedesignedPhasechange2025, delaney-muskens2020family, radfordInverseDesignUnitary2025}, and support in silico inverse design, where a computer-simulated and discovered design is transferred to a PCM device, or in situ design with single-pixel perturbation optimization. However, the update speed and write endurance of PCM platforms present obstacles for gradient-based inverse-design approaches.

Outside the context of inverse design of conventional photonic functions and devices, free-form programmable photonic platforms have also been introduced and used for realizing on-chip optical neural networks: Refs.~\cite{wuLithographyfreeReconfigurableIntegrated2023,wuFieldprogrammablePhotonicNonlinearity2025} are based on a programmable gain medium that allows the modulation of the imaginary part of the complex refractive index, and our prior work~\cite{Onodera2025} is based on a programmable lithium-niobate waveguide using the same operating principle as the device we used in the present work.

In this work we develop and apply techniques for in situ optimization of a free-form programmable device to achieve conventional photonic functions, controlling light in both the spatial and spectral domains. We demonstrate inverse design of a single programmable device to perform a variety of functions. Our work simultaneously (1) mitigates the simulation--reality gap in conventional inverse design, and (2) breaks from the conventional paradigm of each inverse-designed function requiring the fabrication of a new physical device.

The physical platform we used is a lithium-niobate slab waveguide whose refractive-index distribution $\Delta n(x,z)$ can be reconfigured on a timescale of seconds~\cite{Onodera2025}. On-top illumination-controlled electro-optic modulation provides free-form $\Delta n(x,z)$ programmability across roughly $10^4$ regions in a lithium-niobate slab waveguide. Because the waveguide's refractive index is programmable, optimization of the refractive index in inverse design can use experimental data from real-time measurements, mitigating any simulation--reality gap~\cite{wrightDeepPhysicalNeural2022}. Using an in situ inverse-design method, we realized a range of common intensity-domain photonic functions: four-\unitforN{} and eight-\unitforN{} permutations, and the demultiplexing of four Hermite--Gaussian modes, with worst-case crosstalk ranging from \SI{-27}{\decibel} to \SI{-13}{\decibel}, as well as matrix--vector multiplication (MVM) with various random matrices. We further demonstrated engineering of wavelength response, enabling both wavelength-independent functionality throughout the \SIrange{1500}{1600}{\nano\metre} window and wavelength-dependent functionality: wavelength demultiplexing for \SI{1310}{\nano\metre} and \SI{1550}{\nano\metre} with crosstalk of \SI{-8.4}{\decibel}.

\section{Programmable Photonic Platform and Inverse Design Approach}

\par We demonstrate in situ programming of inverse-designed devices with a photoconductor-covered slab waveguide whose refractive-index distribution $\Delta n(x,z)$ is set by projecting an illumination pattern $I(x,z)$ onto the waveguide in operation. The index distribution can be reset in about \SI{1.5}{\second} by turning off the programming illumination, or set to another pattern with a new projection. For any desired photonic functionality, an appropriate refractive-index distribution and its corresponding illumination pattern can be inverse-designed in minutes via back-propagation through a digital model with closed-loop feedback from the physical waveguide.

\subsection{Slab waveguide design and experimental setup}

\par To make a slab waveguide that is susceptible to programming by an illumination pattern, we sandwich a \SI{700}{\nano\metre} thin film lithium niobate (TFLN) core with \SI{1}{\micro\metre} SiO$_2$ claddings on each side on top of a conductive silicon substrate. We then deposit a \SI{6}{\micro\metre}-thick silicon-rich nitride (SRN) photoconductive layer, as well as a \SI{30}{\nano\metre} indium tin oxide (ITO) layer on top of it as an electrode. Details are provided in Appendix~\ref{sec:fabrication}. We used metal probes attached to both the substrate and the top electrode to apply a large AC voltage across the whole stack. The photoconductive layer and the waveguide stack form a voltage divider structure that functions as follows: When the photoconductor is locally illuminated, it becomes more conductive and the applied voltage redistributes vertically to create a strong field across the lithium niobate waveguide.
As a result, the refractive index of the lithium niobate increases via the electro-optic (EO) effect, raising the effective index of the transverse magnetic slab mode TM$_{0}$ in the illuminated region. 
Therefore, by controlling the illumination pattern $I(x,z)$, we effectively control the refractive index distribution in the waveguide. 
The induced refractive index change increases with larger applied voltage, but the voltage needs to remain low enough to avoid dielectric breakdown, while the AC frequency needs to remain within a window that permits the electric fields to permeate the dielectric waveguide layers and the photoconductor. Here, we chose an AC amplitude of \SI{600}{\volt} and a frequency of \SI{4}{\hertz}, resulting in an index change of $1.7\times10^{-3}$. This is a larger index change, but at a slower AC frequency than in our previous work, where speed was paramount for large machine-learning datasets \cite{Onodera2025}. For further theoretical background about the index modulations, see Appendix~\ref{sec:design}; for experimental measurements, see Appendix~\ref{sec:holography}. 

When we update the illumination pattern, the refractive-index distribution reaches a new steady state within \SI{1.5}{\second}.
Illumination pattern are projected onto the top of the device using a digital micromirror device (DMD) illuminated by an LED. We use a maximum intensity of \SI{50}{\milli\watt\per\square\centi\metre} on the chip. To create inputs for the inverse-designed devices, coherent laser light is shaped into a line with controllable amplitude and phase at each position, $E_\text{in}(x) \coloneqq E(x,z=0)$, by a spatial light modulator (SLM) and coupled through polished side facets of the chip. The outputs $E_\text{out}(x) \coloneqq E(x,z=L)$ are inferred by imaging the output intensity distribution $I_\text{out}(x)$ at the output facet of the chip on a camera and taking the pixel-wise square root to find the field amplitude, disregarding phase information. The complete details for the experimental setup are provided in Appendix~\ref{sec:setup}.

\subsection{Digital model}

\par To inverse-design devices, we built an auto-differentiable digital model for wave propagation in the waveguide, ${f: (E_\text{in}, I) \mapsto E_\text{out}}$, that maps a given input optical field $E_\text{in}(x)$ and illumination pattern $I(x,z)$ to an output optical field $E_\text{out}(x)$ and, consequently, intensity $I_\text{out}(x)$. The schematic of the model is shown in Appendix~\ref{sec:DT illustration} and Fig.~\ref{fig:DT illustration}. The digital model is based mainly on the two-dimensional beam-propagation method, solving the equation
\begin{align}
    \label{eq:beam_propagation}
    \pder{E}{z} = \frac{\im}{2k}\pder{^2E}{x^2} + \im \Delta n(x, z) k_0 E.
\end{align}
Where $k_0$ is the wave number for the light. The overall index pattern in the device $\Delta n(x,z)$ comprises two parts, the illumination-induced programmable index change $\Delta n_\text{prog}(x,z)$ and static background index perturbations $\Delta n_\text{bg}(x,z)$. We constructed an empirical physics-based model that estimates $\Delta n_\text{prog}(x,z)$ from $I(x,z)$. The empirical model accounts for a variety of misalignments between the programming illumination and the chip, and it models the nonlinear relation between intensity $I(x,z)$ and illumination-induced index change $\Delta n_\text{prog}(x,z)$.
We also trained a convolutional neural network (CNN) as a correction term to help the empirical physics-based model to process illumination patterns $I(x,z)$ before converting them to refractive-index distributions $\Delta n_\text{prog}(x,z)$, locally correcting for uneven illumination brightness and various imaging imperfections. Additionally, we fitted the background refractive-index perturbations $\Delta n_\text{bg}(x,z)$ to model distortions of the wavefront passing through the device. The origin of said distortions is unclear, and the distortion cannot be modeled by damaged facets or misalignments alone. The total refractive-index perturbation entering Eq.~\eqref{eq:beam_propagation} is then $\Delta n(x,z)= \Delta n_\text{prog}(x,z)+\Delta n_\text{bg}(x,z)$.

\par This empirical model has several unknown parameters, particularly the static background refractive-index perturbation $\Delta n_\text{bg}(x,z)$, the weights of the convolutional neural network, and any parameters modeling misalignments of the experimental system. We optimized these parameters via auto-differentiation to make accurate predictions of the experiment. To provide ground-truth data for this optimization, we built a large dataset consisting of 10{,}000 random illumination patterns and 50 random input fields for each illumination pattern. We generated illumination patterns $I(x,z)$ roughly resembling the final patterns of inverse-designed devices by creating random patterns containing low spatial-frequency features. For the optical input field, we also generated low spatial-frequency noise and random phase jumps. For each pair of illumination pattern and input field, we collected the output intensity distribution at the camera, providing 500{,}000 ground-truth pairs $\{\big((E_\text{in}^{(i)}, I^{(i)}), I_\text{out}^{(i)}\big)\}$. The details of dataset generation are provided in Appendix~\ref{sec:dataset}. We then fit the digital model to the dataset. The details of model fine-tuning and agreements between predictions and measurements are provided in Appendix~\ref{sec:DT fine-tuning}.

\subsection{Inverse design process}
\label{sec:Inverse design process}

\par To inverse-design photonic devices, we define their desired functionality with a loss function and minimize this loss using the physics-aware training algorithm~\cite{wrightDeepPhysicalNeural2022}, as shown in Fig.~\ref{fig:1}. We refer to this approach as \textit{in situ photonic inverse design} because instead of optimizing the loss function entirely in silico as is conventionally done, we use in situ measurements of the fabricated device's actual behavior within the optimization loop. The optimization is performed by alternating between (1) evaluating the loss function based on experimental measurements of the behavior of the photonic chip with the current programmable parameters (illumination pattern $I(x,z)$), and (2) using gradient descent via backpropagation of an in silico (digital) model of the behavior of the photonic chip to determine updates to the programmable parameters. The ability to rapidly reprogram the behavior of the device in situ enables this alternative approach to photonic inverse design. Incorporating measured experimental results in the optimization loop allows the inverse design process to be more tolerant, versus conventional in silico approaches, of any simulation--reality gap between the experiment and the digital model, as might arise from having imperfect calibration or imperfect knowledge of the device physics (see Appendix~\ref{sec:PAT_advantage}).
Since we can only record output intensity in our experimental setup, we made the simplifying assumption that the output phase is flat, and the output field is simply estimated to be the square root of the output intensity. 
This allowed us to train intensity-based functions without phase-imaging capabilities (however, we note that there are no fundamental obstacles to implementing phase-imaging alongside this experimental setup). 
We designed a loss function that rewards power coupled into target output modes and suppresses undesired crosstalk. For details about each loss function used, see Section~\ref{sec:ExperimentalResults} below and Appendix~\ref{sec:PAT}.

\par During one iteration of the optimization loop for an $N$-input, $N$-output ($N\times N$) functionality, in the forward-pass phase of the loop, we measure the intensity output $I_{\text{out}}^{(i)} = |E_{\text{out}}^{(i)}(x)|^2$ of the device for each of $N$ different input fields $E_{\text{in}}^{(i)}(x)$, with $i$ ranging from $1$ to $N$. Each measurement is performed in a single experiment: input the field and measure the output intensity. 
The backward pass involves computing the gradient using the updated loss, which determines the update to the illumination pattern $I(x,z)$. The inverse design converged in 100 iterations for all tasks, which took around 10 to 20~minutes in each case. The longest step in this procedure was the data-collection process for the ground-truth dataset $\{\big((E_\text{in}^{(i)}, I^{(i)}), I_\text{out}^{(i)}\big)\}$, which took around 43~hours. However, once the data is collected and the model is trained, many target tasks can be designed using the same model, and the model can predict outputs accurately for weeks. 

\subsection{Characterizing device performance}

\par To monitor the performance during inverse design, as well as characterize the results after the inverse design was completed, we calculated the crosstalk, insertion losses and matrix fidelity using the recorded output intensity distribution. 
For integrated photonic devices, the insertion loss is usually defined as the ratio of driving power at a given input \unitforN{} to measured power in the target output \unitforN{}, expressed in decibels. 
Since we did not measure input and output power at each \unitforN{} in absolute terms, we estimate the insertion loss by calculating normalized mode overlaps between the estimated output fields and the target fields. Our procedure is detailed in Appendix~\ref{sec:insertion}, but we note here that our results should be interpreted as a lower bound on the actual insertion loss. 
For the tasks demonstrated in this work, the input and target output distributions are Hermite--Gaussian beams or spatially separated Gaussian beams. While these beams do not correspond to eigenmodes of the slab waveguide, the tasks we performed here are similar to inverse-designed devices which perform operations on the modes of channel waveguides~\cite{luInversedesignedUltracompactStarcrossings2017, xu5x5Mm2Compact2017, shangInverseDesignedLithiumNiobate2023, changUltracompactModeMultiplexer2018, xieHighlyCompactEfficient2020, sunEdgeguidedInverseDesign2025, wuCompactScalableMode2025}. Therefore, we still refer to these beams as \textit{\unitforN{}s} for simplicity, although no channel waveguides were fabricated in this work. 
Since future systems would likely couple from channel waveguides into the programmable waveguide, we purposely neglect losses due to free-space-to-chip coupling (both at the input and output facets).
However, our estimates also neglect absorption and radiation losses (which we estimate at roughly \SI{1}{\decibel/\centi\metre}~\cite{Onodera2025}), as well as coupling losses due to suboptimal (non-flat) phase-fronts.
Our estimates for crosstalk are derived similarly, with the only inaccuracy stemming from the simplifying assumption of flat phase-fronts, as detailed in Appendix~\ref{sec:insertion}.

\begin{figure}[!htbp]
    \centering
    \includegraphics[width=1\linewidth]{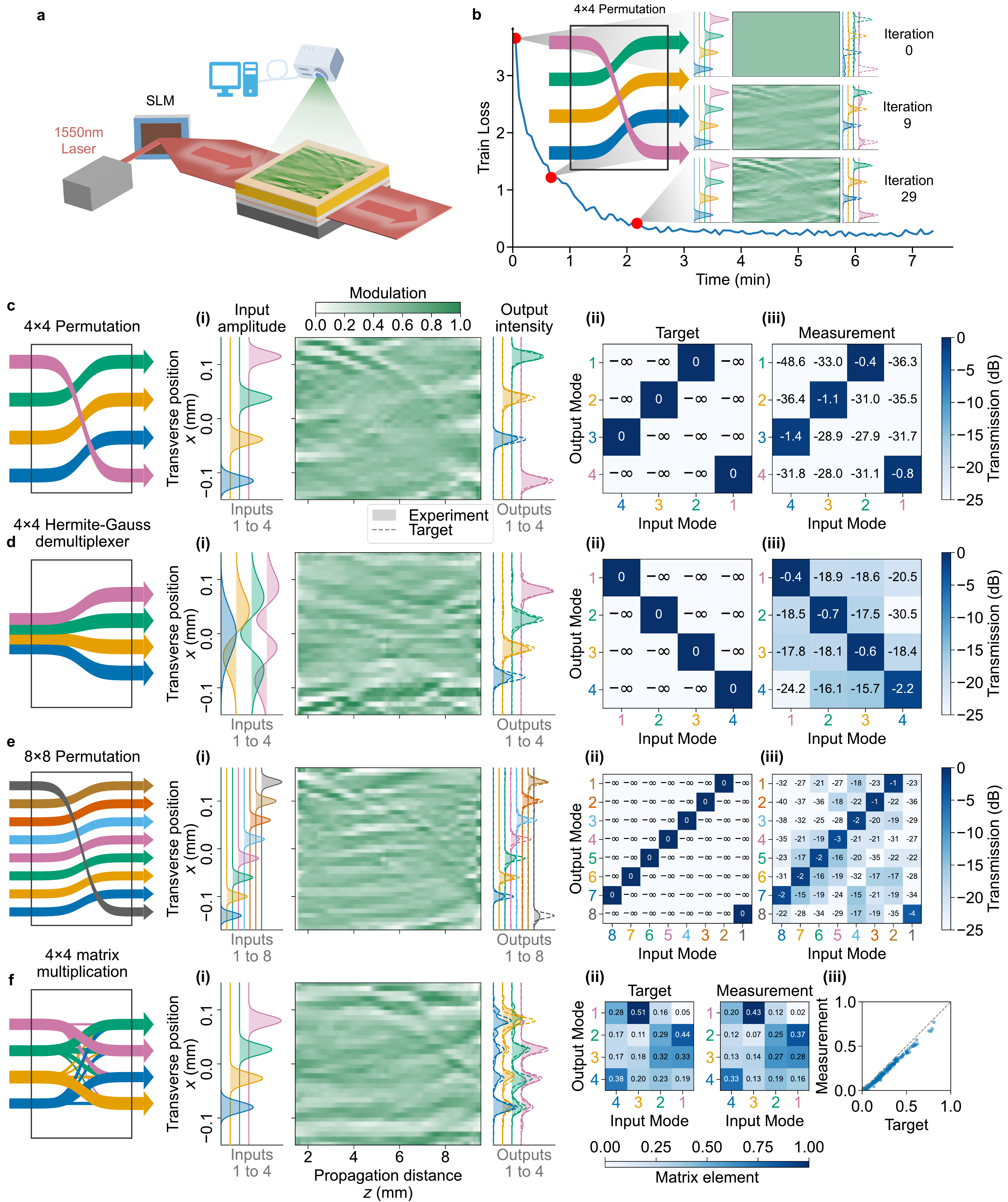}
    \caption{\textbf{Reconfigurable inverse design of typical photonic devices.} \textbf{a}, Illustration of the inverse design experimental setup. SLM: spatial light modulator. \textbf{b}, Training loss for $4\times4$ permutation as a function of training time; the three plots on the side show the patterns and output measurements at different iterations of the optimization loop. \textbf{c}, Inverse-designed $4\times4$ spatial-mode permutation using the physics-aware training (PAT) method. (i)~Input amplitude, inverse-designed illumination pattern, measured output intensity, and target intensity. (ii)~Target transmission matrix. (iii)~Measured transmission matrix. \textbf{d}, Inverse-designed $4\times4$ Hermite--Gaussian mode demultiplexer and its performance. \textbf{e}, Inverse-designed $8\times8$ spatial-mode permutation and its performance. \textbf{f}, Inverse-designed $4\times4$ random matrix and its performance. (i)~Input amplitude, inverse-designed illumination pattern, measured output intensity, and target intensity. (ii)~Target and measured intensity transmission matrices. (iii)~Measured vs. target matrix elements for 10 different random matrices.}
    \label{fig:2}
\end{figure}

\section{Experimental Results}
\label{sec:ExperimentalResults}

\subsection{Inverse-Designed Single-Wavelength Devices}
\label{sec:Single_wavelength}

\par We first present canonical devices commonly demonstrated in inverse design. We demonstrate a $4\times4$ spatial Gaussian-mode permutation, a $4\times4$ Hermite--Gaussian mode demultiplexer, an $8\times8$ spatial Gaussian-mode permutation, and $4\times4$ random matrix multiplication. 
Note that all demonstrations were only characterized using intensity measurements, since we could only record output intensities on our camera, and we assumed a flat output phase for all the results. We used coherent monochromatic \SI{1550}{\nano\metre} light shaped by a 1D beam shaper to probe the inverse-designed devices in this section (Fig.~\ref{fig:2}a). We used the free-space beam shaper to create each input \unitforN{} one at a time, collected the corresponding output intensity, and divided it by the average power of all the \unitforN{}s, normalizing them while preserving the relative power differences. We then followed the PAT algorithm using a loss function $\mathcal{L}_\lambda$ consisting of mode overlaps between the measured and target outputs, detailed in Appendix~\ref{sec:PAT}.
We recorded the loss function value in real time as a function of how many iterations of the optimization loop had been completed, shown in Fig.~\ref{fig:2}b for the $4\times4$ permutation task as an example. The training converged in around 7 minutes as we observed the illumination pattern changing and output intensity distribution improving relative to the target in real time.

\par We achieved \SI{-27}{\decibel} crosstalk for the $4\times4$ spatial Gaussian-mode permutation task (Fig.~\ref{fig:2}c), \SI{-13}{\decibel} crosstalk for the $4\times4$ Hermite--Gaussian mode demultiplexing (Fig.~\ref{fig:2}d) and \SI{-13}{\decibel} crosstalk for the $8\times8$ spatial Gaussian-mode permutation (Fig.~\ref{fig:2}e), with performance similar to previous work using conventional inverse design. A more detailed comparison is provided in Table~\ref{tab:performance_comparison_lit}. The crosstalk arises from several sources. 
First, even a device that behaves perfectly according to the wave equation (Eq.~\ref{eq:beam_propagation}), with the same maximal $\Delta n_\text{prog}$ and the same length as our device, only achieves insertion losses and crosstalk a few \SI{}{\decibel} lower than the experimental demonstrations presented here.
Second, our digital model may still fail to perfectly estimate gradients. This could be mitigated by improving the model further, but some imperfections of the device, e.g. those modeled by $\Delta n_\text{bg}$, may also introduce uncorrectable perturbations, for example due to features that are smaller than the programmable pixel size. 
Finally, noise from the measurement setup, especially an uneven and fluctuating noise floor on our IR camera, distorts the output measurement. Nevertheless, the ability to programmably control the device function can largely enhance its utility (see Appendix~\ref{sec:PAT_advantage}).

\par We also demonstrated matrix--vector multiplication of vectors encoded in the intensity of Gaussian beams with $4\times4$ random matrices (Fig.~\ref{fig:2}f). 
Although the device manipulates coherent optical fields governed by $E_{\mathrm{out},i} = \sum_jU_{ij}E_{\mathrm{in},j}$, demonstrating linear matrix--vector multiplication directly using only intensity-measurements is complicated by interference among multiple input \unitforN{}s, destroying the linear relation between input and output intensities. Instead, we send in basis vectors only, launching each \unitforN{} sequentially, to effectively perform incoherent intensity matrix multiplication free of interference between \unitforN{}s. Under this scheme, the system acts as a linear operator on the input intensities, $I_{\mathrm{out},i} = \sum_j|U_{ij}|^2I_{\mathrm{in},j}$, where each matrix element is determined by the fraction of optical power coupled from a given input into a target output \unitforN{}. We optimize this power-transfer mapping using a normalized mean-squared error (NMSE) objective~\cite{paiMatrixOptimizationUniversal2019a}.
We collected measurements for ten different randomly sampled matrices and show the measurements for a representative matrix in Fig.~\ref{fig:2}f(ii), and the correlation between target and measured matrix elements for all ten matrices in Fig.~\ref{fig:2}f(iii).

\par The trained patterns operated reliably over a considerable time span. We tested the same pattern for the $4\times4$ permutation task across a time window of 2 weeks, and the performance remained mostly the same. We tested by applying the pattern and measuring the output, but we note that the device should be left in uniform illumination or in the dark when not in use. Details are provided in Appendix~\ref{sec:Endurance}.

\begin{figure}[!htbp]
    \centering
    \includegraphics[width=1\linewidth]{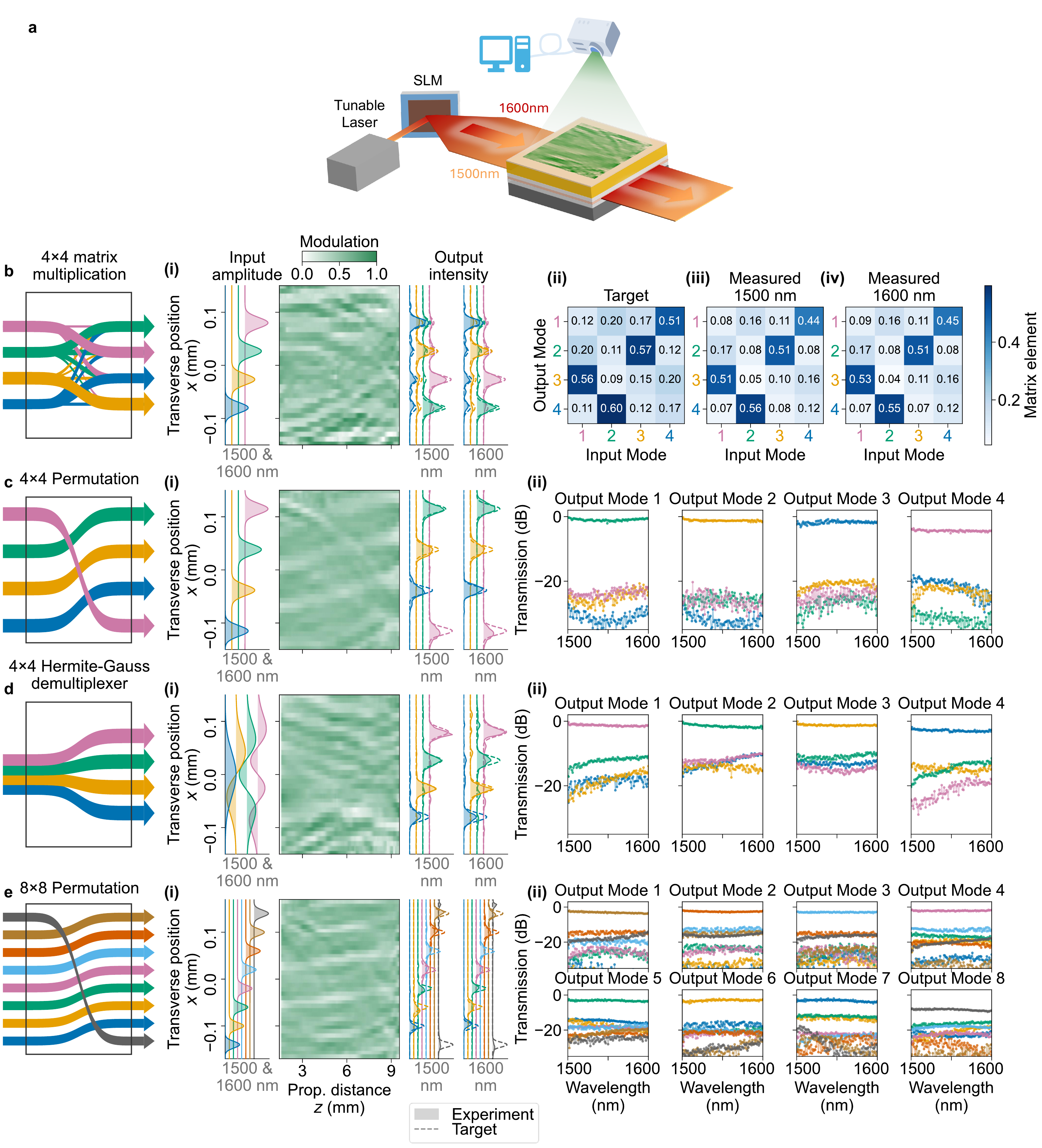}
    \caption{
    \textbf{Broadband inverse design of typical photonic devices.}
    \textbf{a}, Illustration of the inverse design experimental setup featuring a \SI{1500}{\nano\metre} to \SI{1600}{\nano\metre} tunable laser. The inverse design process combines the loss functions measured at \SI{1500}{\nano\metre} and \SI{1600}{\nano\metre} and optimizes them simultaneously. SLM: spatial light modulator.
    \textbf{b}, Inverse-designed broadband $4\times4$ random matrix and its performance. (i)~Input amplitude, inverse-designed illumination pattern, measured output intensity, and target intensity at \SI{1500}{\nano\metre} and \SI{1600}{\nano\metre}. (ii)~Target matrix. (iii)~Measured matrix at \SI{1500}{\nano\metre}. (iv)~Measured matrix at \SI{1600}{\nano\metre}.
    \textbf{c}, Inverse-designed broadband $4\times4$ spatial-mode permutation using the physics-aware training (PAT) method. (i)~Input amplitude, inverse-designed illumination pattern, measured output intensity, and target intensity at \SI{1500}{\nano\metre} and \SI{1600}{\nano\metre}. (ii)~Transmission for each output \unitforN{} from each input \unitforN{} over the spectrum from \SI{1500}{\nano\metre} to \SI{1600}{\nano\metre}.
    \textbf{d}, Inverse-designed broadband $4\times4$ Hermite--Gaussian mode demultiplexer and its performance.
    \textbf{e}, Inverse-designed broadband $8\times8$ spatial-mode permutation and its performance.
    }
    \label{fig:3}
\end{figure}

\subsection{Inverse-Designed Broadband Devices}
\label{sec:Main_broadband}

\par Just as we can engineer spatial-domain functionalities, we can engineer spectral behavior using in situ inverse design. Here we demonstrate that we can reduce the wavelength dependence of output response across a large part of the telecom spectrum to produce devices that are equalized from \SIrange{1500}{1600}{\nano\metre} in minutes.
Our laser has a tuning range from \SIrange{1500}{1630}{\nano\metre} (Fig.~\ref{fig:3}a) and we modified the loss function for inverse design to include the loss at both \SI{1500}{\nano\metre} and \SI{1600}{\nano\metre}, while using the same digital model for PAT that was trained at \SI{1550}{\nano\metre}. Details on how we performed free-space beam shaping at multiple wavelengths and modeled waveguide dispersion are provided in Appendix~\ref{sec:multilam}. We first demonstrated that we could design a refractive index pattern that can implement the same matrix multiplication at \SI{1500}{\nano\metre} and \SI{1600}{\nano\metre} (Fig.~\ref{fig:3}b(ii)-(iv)). We also demonstrated the same canonical inverse-design tasks from Section~\ref{sec:Single_wavelength} and measured their transmission matrices across the \SIrange{1500}{1600}{\nano\metre} range at \SI{1}{\nano\metre} resolution (Fig.~\ref{fig:3}c--e). 
We note that the performance is slightly worse for broadband demonstrations compared to single-wavelength performance, presumably due to the digital model having been trained at primarily one wavelength and the device not being able to completely equalize the performance at different wavelengths.
Nonetheless, such in situ design of broadband devices could enable wavelength-independent manipulation of spatial modes across a large bandwidth, leading to high-bandwidth optical parallel computing and wavelength-noise-resilient devices.

\subsection{Inverse-Designed Wavelength Demultiplexer}
\label{sec:Main_WDM}
We also show the opposite behavior, namely that we can design refractive index patterns to produce wavelength-dependent behavior and utilize this to separate signals in the same spatial \unitforN{} at different wavelengths. We demonstrate a two-wavelength demultiplexer between \SI{1310}{\nano\metre} and \SI{1550}{\nano\metre} using in situ inverse design. 

\begin{figure}[!htbp]
    \centering
    \includegraphics[width=1\linewidth]{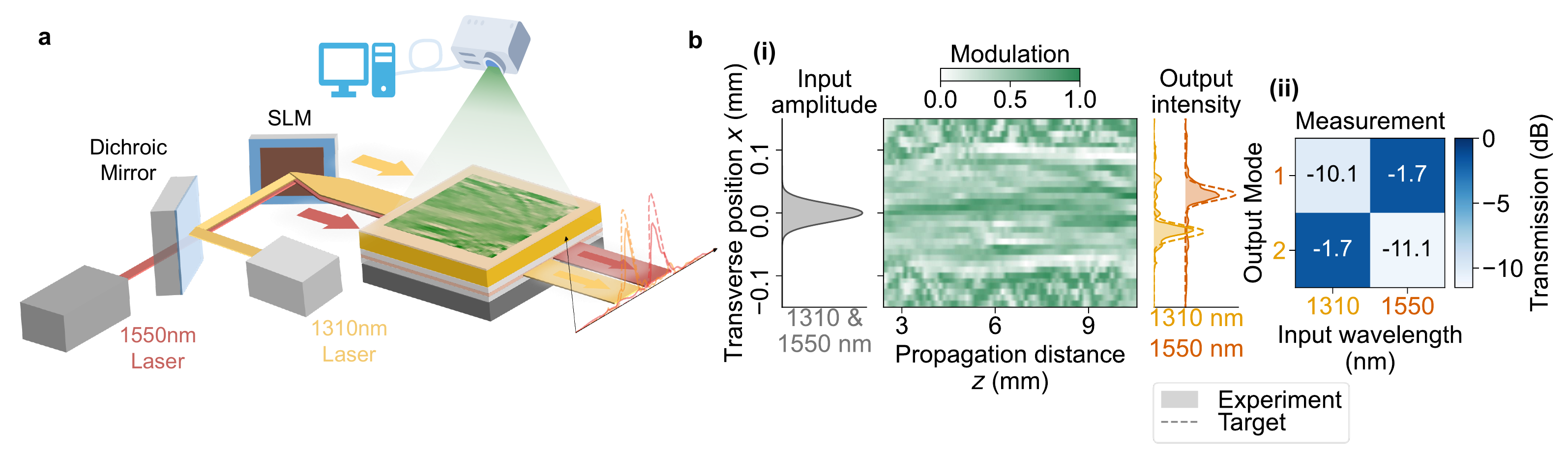}
    \caption{\textbf{Inverse-designed wavelength demultiplexer.} \textbf{a}, Illustration of the wavelength-multiplexing inverse design experimental setup with a dichroic mirror combining the \SI{1550}{\nano\metre} and \SI{1310}{\nano\metre} lasers. SLM: spatial light modulator. \textbf{b}, Inverse-designed two-wavelength demultiplexer and its performance. (i)~Input amplitude, inverse-designed illumination pattern, measured output intensity, and target intensity at \SI{1310}{\nano\metre} and \SI{1550}{\nano\metre}. (ii)~Transmission matrix.}
    \label{fig:4}
\end{figure}

To accurately predict gradients for the refractive index distribution at \SI{1310}{\nano\metre}, we retrained the digital model of the experiment on a new, task-specific illumination pattern dataset, for which we created 10{,}000 illumination patterns from many wavelength-division demultiplexers based on perturbed background index map and trained entirely in silico. We double the dataset size by taking the average of 5 randomly picked patterns, resulting in a total of 20{,}000 patterns. We collected experimental input-output pairs for these illumination patterns for inputs using only \SI{1550}{\nano\metre} light. We then trained the digital model to accurately predict the input-output pairs at \SI{1550}{\nano\metre}. During physics-aware training, we added another \SI{1310}{\nano\metre} laser (DFL Series, SS63310100007, Lasertel) and combined it with the \SI{1550}{\nano\metre} laser using a dichroic mirror (DMLP1500, Thorlabs) (Fig.~\ref{fig:4}a). During each forward-pass phase of the optimization loop, we sent in the spatial \unitforN{} at two wavelengths sequentially, and recorded the corresponding output intensities. We used a loss function summing contributions from both wavelengths, and optimizing the programmed pattern to satisfy the targets at both wavelengths simultaneously.

The results are shown in Fig.~\ref{fig:4}b, demonstrating the inverse-designed wavelength demultiplexer with a crosstalk of \SI{-8.4}{\decibel}.

\section{Discussion}
\label{sec:discussion}

\subsection{Summary of the Work}

We used a two-dimensionally programmable lithium-niobate waveguide to demonstrate inverse design of various photonic functions via in situ optimization. During optimization, the waveguide's programmable refractive-index pattern was updated incorporating real-time measurements of the device, compensating for fabrication and calibration imperfections and simulation inaccuracies.

Using this method, we programmed a single device to perform a wide range of photonic functions: $4\times4$ and $8\times8$ Gaussian-mode permutations, $4\times4$ Hermite--Gaussian mode demultiplexing, and random matrix--vector multiplication. The maximum crosstalk was \SI{-27}{\decibel} for the $4\times4$ permutation, \SI{-13}{\decibel} for the $8\times8$ permutation, and \SI{-13}{\decibel} for Hermite--Gaussian demultiplexing. A full comparison with single-function devices that were produced using conventional inverse design and fabrication is shown in Table~\ref{tab:performance_comparison_lit}. We also demonstrated control over our device's spectral response. We demonstrated wavelength-independent operation of the same four- and eight-\unitforN{} transformations across the \SIrange{1500}{1600}{\nano\metre} window, and wavelength demultiplexing between \SI{1310}{\nano\metre} and \SI{1550}{\nano\metre} with maximum crosstalk of \SI{-8.4}{\decibel}. All functions were designed and implemented on the same device by changing only the projected illumination pattern. 
\subsection{Limitations}

The combination of programmable wave propagation and inverse design could enable deployment of multimode photonic devices in scenarios where the development of application-specific devices is uneconomical, as well as in scenarios where it is useful for a single device to switch between functions. The prototype reported here is a laboratory test device that is not yet suitable for widespread use. In this section, we discuss what key improvements would advance the device platform we have used towards broader utility.

The principal physical limitation is the small programmable index contrast of approximately $1.7\times10^{-3}$, 
limiting the compactness of the designs and the number of input and output \unitforN{}s of our demonstrations. There are at least three approaches to increasing the index contrast: improving the photoconductor switching ratio to increase the electric-field contrast between the bright and dark states, which we estimate would yield an index change as large as $10^{-2}$ (Appendix~\ref{sec: Future Device}); restructuring the device as a three-layer stack with the photoconductor serving as waveguide cladding (Appendix~\ref{sec:DC device}); or replacing the waveguide core with a material of higher electro-optic coefficient, such as an electro-optic polymer (Appendix~\ref{sec: Polymer Device}). At an index change of $10^{-2}$, simulations show that $128\times128$ spatial permutations and $128\times128$ mode demultiplexing become achievable with a chip footprint of $\SI{5}{\centi\metre}\times\SI{5}{\centi\metre}$ (Appendix~\ref{sec:large scale}).

A second limitation is the coarse spatial resolution of the programmable index, limited in part by the \SI{9}{\micro\metre} pitch of the projected digital-micromirror-device pixels. The large pixels produce a smooth index profile, which is an advantage for broadband operation but a disadvantage for wavelength selectivity. Both the projection optics and the lateral spreading of the programming field across device layers contribute to the present resolution limit, which is ultimately bounded by the latter. The device geometries discussed in Appendices~\ref{sec: Future Device} and~\ref{sec:DC device} could address this field-fringing issue.

A third limitation is that the programmed index pattern persists only while the illumination and the bias are applied. In the present device, the static power consumption has two sources: approximately \SI{3}{\milli\watt} of optical power for the programming illumination, and approximately \SI{2}{\milli\watt} of electrical power for the bias. Improvements to the photoconductor switching ratio and to the device stack (Appendix~\ref{sec: Future Device}) should reduce the illumination requirement to the \si{\micro\watt} level while simultaneously increasing operating speed, and quasi-DC operation (Appendix~\ref{sec:flickering}) should bring the electrical power to the \si{\micro\watt} level as well. Ultimately, programming non-volatile refractive-index changes would remove static illumination energy consumption entirely and only leave static electrical energy consumption, while remaining rewritable. Such devices may be attainable by using programmable ferroelectric-domain poling of the TFLN waveguide core (Appendix~\ref{sec: Non-volatile}).

A fourth limitation is speed. There are two types of speed for the device and experiments we reported: the input--output recording speed and the pattern switching speed. Output measurements were synchronized to the \SI{4}{\hertz} AC bias, and a change of illumination pattern required approximately \SI{1.5}{\second} to reach steady state. A faster measurement setup could increase the recording speed to the \si{\giga\hertz} level, and a higher-conductivity photoconductor combined with quasi-DC operation (Appendix~\ref{sec:flickering}) should raise the pattern switching speed to the \si{\kilo\hertz} level (Appendix~\ref{sec:performance}). For applications like adaptive calibration, loading new weights into an optical neural network, or reconfiguring a mode-routing function as network demand changes, the projected kilohertz operation should be ample. More demanding applications like optical circuit switching within data centers require switching times on the order of \SI{10}{\micro\second} to rival electronic packet routing in that setting~\cite{satoOpticalSwitchingWill2024}. 
The projected pattern switching speed is limited mostly by the switching speed of the photoconductor; alternative photoconductive materials could further increase the switching speed.

A fifth limitation is a weak residual memory that occurs after prolonged exposure to a fixed non-uniform illumination pattern: we observed that a residual index pattern remained for hours after the programming illumination was removed. The effect was reversible, with uniform illumination erasing it and leaving no permanent damage. Its physical origin remains unknown and requires further material study (Appendix~\ref{sec:Endurance}). We have only observed this effect with the current prototype device; other devices fabricated with different recipes did not exhibit this effect, so the effect could be addressed by modifying the fabrication recipe.

Further limitations of our experimental results are due to the measurement and operating setup. We recorded only the intensity of the output light, so the output profiles carry a phase ambiguity. Therefore, insertion-loss measurements need to be interpreted as lower bounds on the actual insertion losses (see details in Appendix~\ref{sec:insertion}). Phase measurement techniques like off-axis holography are compatible with our approach and could be implemented in the future. Another limitation is that our device was free-space coupled in our experimental setup, whereas for many applications a fully integrated solution may be preferred. We have proposed a fully integrated version of our device in previous work~\cite{Onodera2025}.

Appendix~\ref{sec:performance} summarizes the device performance, limitations, and potential future improvements, and Table~\ref{tab:performance_comparison} collects the key metrics for current and projected devices.

\subsection{Outlook}
If the device improvements suggested in the above section were realized, devices based on completely unstructured slab waveguides could compactly and efficiently realize a variety of useful photonic functions. However, we could also combine our programmability with existing photonic components, similar to the idea explored in Ref.~\cite{masonHighPerformanceWavelengthDivision2025}, where inverse design was co-optimized with Bragg gratings to minimize crosstalk. We could hand-design a structure to have etched gratings and use real-time programmability to engineer custom spectral responses. For example, we could pre-fabricate Bragg gratings with different periods; then on top of spectral channels with wide spacing, programmability could provide fine tuning of the channels as well as spatial control. There also exists the possibility of adding programmability to fabricated structures---either human-designed or inverse-designed---so that the behavior of the structures could be modified after fabrication using our in situ optimization approach. This could extend from providing a mechanism for trimming---recovering the intended device behavior by allowing for small adjustments to compensate for fabrication imperfections or other effects that cause small simulation--reality gaps---to enabling completely new functions to be realized with some etched structures as an initial canvas. Our approach is compatible with etched structures provided the structures exhibit a usable electro-optic effect and a photoconductor can be deposited on top of them.

Inverse design of many-input, many-output devices is very challenging, in part because the simulation--reality gap grows with device scale and function complexity~\cite{schulInverseDesignScalable2026a}. While simulation and fabrication techniques are continually being pushed to be more accurate, it seems likely that for the foreseeable future there will be many photonic functions where the simulation--reality gap is simply too large for conventional inverse design to succeed. While other works have investigated methods to lower the simulation burden~\cite{nikkhahInversedesignedLowindexcontrastStructures2024} and more reliably predict fabrication outcomes~\cite{grinbergInverseDesignRobust2023}, our approach could potentially enable scaling to device sizes that are impractical for conventional inverse design because of the ability to tolerate and mitigate simulation--reality gaps through programmability. Simultaneously, our approach allows devices whose function can be rapidly changed after deployment.

\section*{Data and code availability}

Experimental data and scripts to replicate the figures in this paper are available at \url{https://doi.org/10.5281/zenodo.22293276}.

\section*{Acknowledgements}
\par We thank NTT Research for their financial and technical support. We gratefully acknowledge the Air Force Office of Scientific Research for funding under Award Number FA9550-22-1-0378, the National Science Foundation for funding under Award Number CCF-1918549, and the Department of Energy Office of Science for funding under Award Number DE-SC0026235. This work was performed in part at the Cornell NanoScale Facility, a member of the National Nanotechnology Coordinated Infrastructure (NNCI), which is supported by the National Science Foundation (Grant NNCI-2025233). P.L.M. acknowledges financial support from a David and Lucile Packard Foundation Fellowship. We acknowledge helpful discussions with Marc Jankowski, Vladimir Kremenetski, Chirag Vaswani, and Ruomin Zhu.

\section*{Author contributions}
T.O., L.G.W., and P.L.M. conceived the project. Y.Z., M.M. Stein, T.O., and P.L.M. designed the devices and experiments. Y.Z., M.M. Stein, B.A.A., and R.Y. performed the device fabrication. T.O. and M.M. Stein built the high-voltage and beam-shaper setups. Y.Z., M.M. Stein, F.O.W. and M.M. Sohoni performed the experiments, and analyzed the results. R.Y. conceived the technique to realize quasi-DC operations with dynamical modulations of the programming illumination, presented in Appendix Section~\ref{sec:flickering}. Y.Z., M.M. Stein, and P.L.M. wrote the manuscript with input from all authors. P.L.M. supervised the project.

\section*{Competing interests}
T.O., M.M.Stein, L.G.W., and P.L.M. are listed as inventors on a patent application (US20250314915A1) on the 2D-programmable waveguide. The other authors declare no competing interests.

\bibliography{references}

\appendix

\section{Device Design and Fabrication}
\subsection{Device Design}
\label{sec:design}
\par
\begin{figure}[!htbp]
    \centering
    \includegraphics[width=1\linewidth]{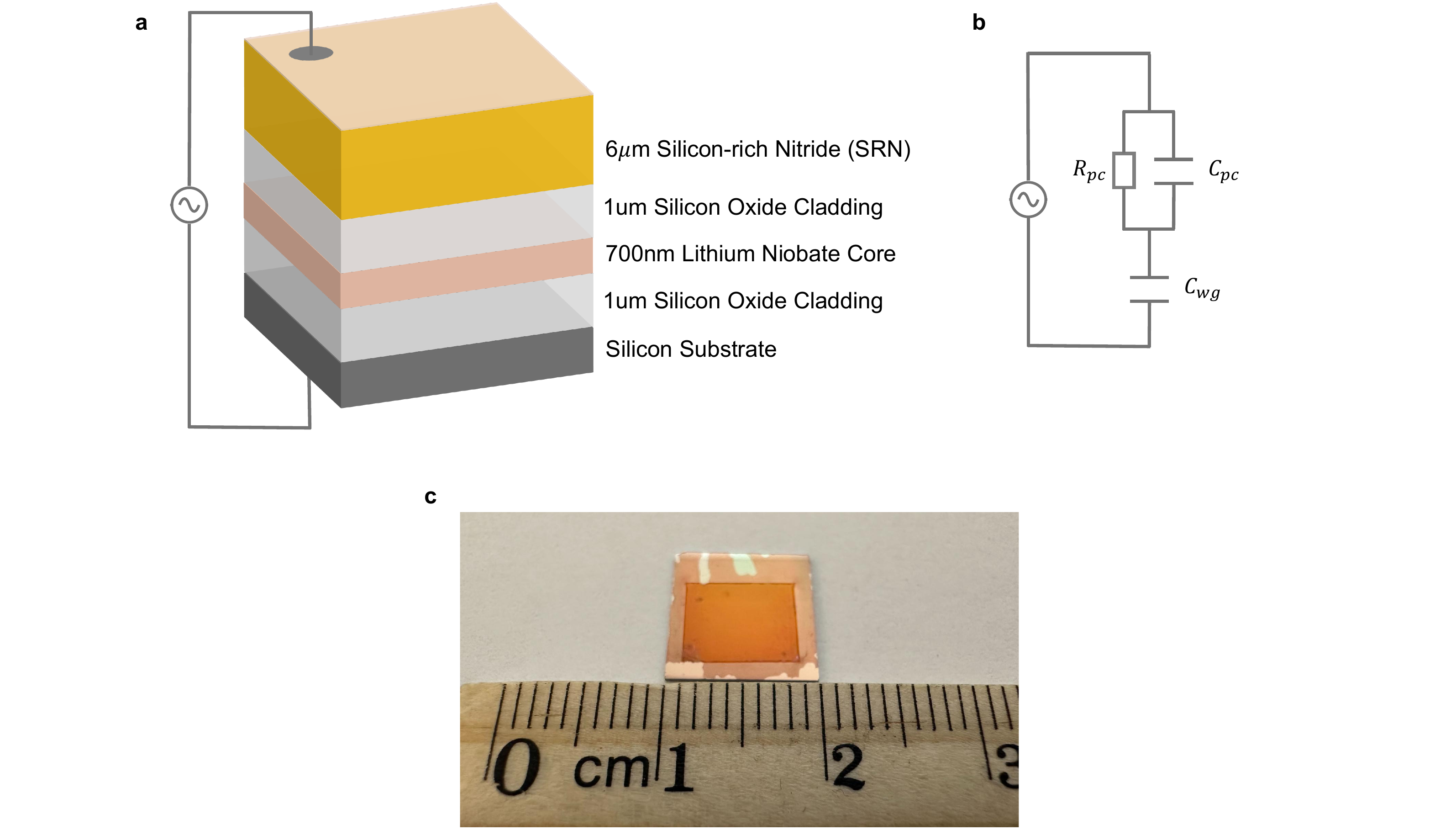}
    \caption{\textbf{Schematics of the device in this work.} \textbf{a}, Cross-section view of the device showing the materials of each layer. The device is driven by an AC voltage of up to \SI{600}{\volt}. \textbf{b}, Simplified electric-circuit model describing the electrical properties and mechanism of the device. \textbf{c}, Photograph of the programmable inverse design device used in this work. The orange rectangle on the chip is the surface of the ITO electrode.}
    \label{fig:S1}
\end{figure}

 Our programmable device is similar in structure to that of our previous work~\cite{Onodera2025}. A schematic of the device is shown in Fig.~\ref{fig:S1}. We used lithium niobate (LN) as the core because it has a large electro-optic coefficient and supports low-loss waveguides~\cite{zhuIntegratedPhotonicsThinfilm2021}. High-quality thin-film LN wafers are also commercially available. We chose z-cut LN so that the bias field in our device is along the crystal $z$-axis, achieving the strongest electro-optic response in a lithium niobate crystal by activating $r_{33}$. This requires the optical field to be aligned with the externally applied bias field, both along the optic axis, which is the $y$ axis in our waveguide coordinates ($x$: in-plane transverse, $y$: out-of-plane, $z$: propagation). We therefore designed the waveguide to support a transverse-magnetic (TM)-polarized mode whose dominant electric-field component is along the $y$ axis.

\par We chose silicon dioxide as the cladding material. Silicon dioxide is the standard bottom cladding in commercial thin-film LN wafers and can also be deposited by plasma-enhanced chemical vapor deposition (PECVD) as the upper cladding. Silicon dioxide offers low absorption and a large index contrast with LN~\cite{zhuIntegratedPhotonicsThinfilm2021}. We chose a \SI{700}{\nano\metre} z-cut lithium niobate film and a \SI{1}{\micro\metre} silicon dioxide cladding, resulting in only one low-loss TM mode supported by the slab, which also has a large overlap with the lithium niobate core to give a strong electro-optic response.

We chose silicon-rich nitride (SRN) as the photoconductive layer because it combines a moderate switching ratio with a high breakdown field~\cite{Onodera2022CNFSRN}. Its electrical and optical properties can be tuned via the Si:N ratio during deposition. We optimized the electrical parameters of the stack as well as the frequency of the applied bias in order to have a large electric field contrast between the bright and dark states in the lithium niobate core, maximizing the programmable index change. The key parameter is the photoconductor resistance $R_{\mathrm{pc}}$, which should change strongly between the bright and dark states. The photoconductor and the waveguide stack form a voltage divider, so the illumination controls the field in the LN core. We modeled the stack as an equivalent circuit to determine the optimal layer parameters and drive frequency. The chip can be described as a photoconductor layer connected to a waveguide stack in series, and the stack can be treated as a pure capacitance $C_{\mathrm{wg}}$, whereas the photoconductive layer is described by a parallel resistor--capacitor element. The photoconductor admittance is therefore
\begin{equation}
Y_{\mathrm{pc}}=\frac{1}{R_{\mathrm{pc}}}+j\omega C_{\mathrm{pc}},
\end{equation}
while the waveguide-stack admittance is
\begin{equation}
Y_{\mathrm{wg}}=j\omega C_{\mathrm{wg}}.
\end{equation}
Under an applied voltage $V_{\mathrm{app}}$, the voltage developed across the waveguide stack follows directly from the series voltage division between these two elements, yielding
\begin{equation}
V_{\mathrm{wg}}
=
V_{\mathrm{app}}
\frac{Y_{\mathrm{pc}}}{Y_{\mathrm{pc}}+Y_{\mathrm{wg}}}.
\end{equation}
Introducing the photoconductor conductance
\begin{equation}
G_{\mathrm{pc}}=\frac{1}{R_{\mathrm{pc}}},
\end{equation}
this may be written as
\begin{equation}
V_{\mathrm{wg}}
=
V_{\mathrm{app}}
\frac{G_{\mathrm{pc}}+j\omega C_{\mathrm{pc}}}
{G_{\mathrm{pc}}+j\omega(C_{\mathrm{pc}}+C_{\mathrm{wg}})}.
\end{equation}
We assume that the index shift scales linearly with the voltage across the waveguide stack,
\begin{equation}
\Delta n = K V_{\mathrm{wg}},
\end{equation}
where $K$ incorporates the electro-optic coefficient, modal overlap, and field-distribution factors.
For dark and bright photoconductive states, characterized by
\begin{equation}
G_{\mathrm d}=\frac{1}{R_{\mathrm d}},\qquad
G_{\mathrm b}=\frac{1}{R_{\mathrm b}},
\end{equation}
respectively, we assume that the photoconductor capacitance remains unchanged upon illumination and define 
\begin{equation}
H_s(\omega)
=
\frac{G_s+j\omega C_{\mathrm{pc}}}
{G_s+j\omega C_\Sigma},
\qquad
s\in\{\mathrm d,\mathrm b\},
\end{equation}
where $C_\Sigma=C_{\mathrm{pc}}+C_{\mathrm{wg}}$. The corresponding complex index-modulation phasor is $\widetilde{\Delta n}_s=K\widetilde V_{\mathrm{app}}H_s$. Because the bright and dark regions experience the same applied AC waveform, the programmable index contrast must be evaluated at the same phase of the drive waveform, i.e., as the difference of the complex phasors:

\begin{align}
\widetilde{\Delta n}_{\text{prog}}(\omega)
&=
K\widetilde V_{\mathrm{app}}
\left[H_{\mathrm b}(\omega)-H_{\mathrm d}(\omega)\right] \\
&=
K\widetilde V_{\mathrm{app}}
\frac{
j\omega C_{\mathrm{wg}}\left(G_{\mathrm b}-G_{\mathrm d}\right)
}{
\left(G_{\mathrm b}+j\omega C_\Sigma\right)
\left(G_{\mathrm d}+j\omega C_\Sigma\right)
}.
\end{align}
Taking the magnitude gives the maximum bright--dark index contrast over an AC cycle,
\begin{equation}
\Delta n_{\text{prog}}(\omega)
=
|K|\,|V_{\mathrm{app}}|
\frac{
\omega C_{\mathrm{wg}}|G_{\mathrm b}-G_{\mathrm d}|
}{
\sqrt{
\left(G_{\mathrm b}^{2}+\omega^{2}C_\Sigma^{2}\right)
\left(G_{\mathrm d}^{2}+\omega^{2}C_\Sigma^{2}\right)
}
}.
\end{equation}

\begin{figure}[!htbp]
    \centering
    \includegraphics[width=1\linewidth]{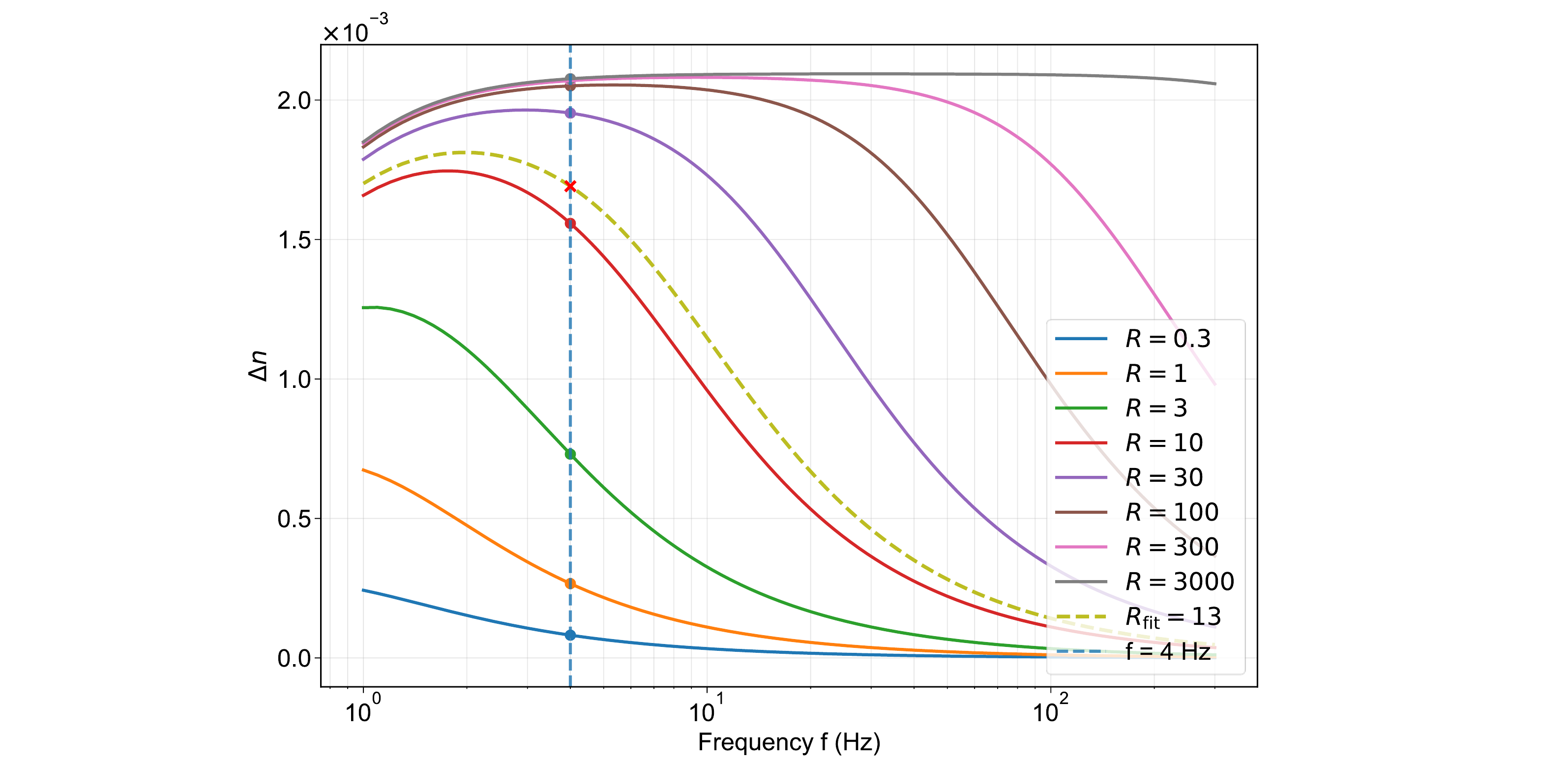}
    \caption{\textbf{Predicted programmable index change versus drive frequency and switching ratio $\mathcal{R}$.} The yellow dashed line shows the frequency curve of the device in this work, and the red cross marks the maximum programmable index change achieved in this work. The electrical parameters for all the curves, and the switching ratio for the yellow dashed curve, were obtained by fitting the measured frequency response in Fig.~\ref{fig:Delta n scan}d.}
    \label{fig:delta n vs f}
\end{figure}

Illumination increases the photoconductor conductance, so $G_{\mathrm b}>G_{\mathrm d}$. Because the conductivity of SRN depends strongly on the applied field~\cite{Onodera2022CNFSRN}, $G_{\mathrm b}$ and $G_{\mathrm d}$ must be evaluated at the voltage that drops across the photoconductor in the bright and dark states, respectively.
Defining the photoconductive switching ratio as
\begin{equation}
\mathcal R
=
\frac{G_{\mathrm b}-G_{\mathrm d}}{G_{\mathrm d}},
\qquad
G_{\mathrm b}=(1+\mathcal R)G_{\mathrm d},
\end{equation}
the programmable index modulation can be written as
\begin{equation}
\Delta n_{\text{prog}}(\omega)
=
|K|\,|V_{\mathrm{app}}|
\frac{
\mathcal R\left(\dfrac{\omega C_{\mathrm{wg}}}{G_{\mathrm d}}\right)
}{
\sqrt{
\left[
(1+\mathcal R)^2+\left(\dfrac{\omega C_\Sigma}{G_{\mathrm d}}\right)^2
\right]
\left[
1+\left(\dfrac{\omega C_\Sigma}{G_{\mathrm d}}\right)^2
\right]
}
}.
\end{equation}
This expression vanishes in both the low- and high-frequency limits and reaches its maximum at
\begin{equation}
\omega_{\mathrm{opt}}
=
\frac{\sqrt{G_{\mathrm b}G_{\mathrm d}}}{C_\Sigma}.
\label{eq:opt_freq}
\end{equation}
At this frequency, the maximum programmable index contrast is
\begin{align}
\Delta n_{\mathrm{prog,max}}
&=
|K|\,|V_{\mathrm{app}}|
\frac{C_{\mathrm{wg}}}{C_\Sigma}
\frac{G_{\mathrm b}-G_{\mathrm d}}{G_{\mathrm b}+G_{\mathrm d}} \\
&=
|K|\,|V_{\mathrm{app}}|
\frac{C_{\mathrm{wg}}}{C_\Sigma}
\frac{\mathcal R}{\mathcal R+2}.
\label{eq:delta_n_max}
\end{align}
Equation~\ref{eq:opt_freq} shows that $\omega_{\mathrm{opt}}$ increases approximately linearly with $G_{\mathrm{pc}}$ at fixed $\mathcal{R}$. Equation~\ref{eq:delta_n_max} shows that $\Delta n_{\mathrm{prog,max}}$ depends only on $\mathcal{R}$ and on the ratio of capacitances $C_{\mathrm{wg}}/C_\Sigma$. We fitted the electrical properties of the device using the measured index change in Fig.~\ref{fig:Delta n scan}. From the fitting, we obtain $K=\SI{6.9}{\per\mega\volt}$, $V_{\mathrm{app}}=\SI{600}{\volt}$, $C_{\mathrm{pc}}=\SI{1.61}{\nano\farad}$, $C_{\mathrm{wg}}=\SI{1.65}{\nano\farad}$, $G_{\mathrm{d}}=\SI{1.09e-8}{\siemens}$ and $G_{\mathrm{b}}=\SI{1.50e-7}{\siemens}$, corresponding to a switching ratio of $\mathcal{R}=12.8$. Representative curves calculated with the fitted parameters of this device are shown in Fig.~\ref{fig:delta n vs f}. Within this idealized RC model, increasing the conductances at fixed $\mathcal R$ does not change $\Delta n_{\mathrm{prog,max}}$. Increasing $\mathcal R$ raises the maximum contrast, with the improvement saturating as $\mathcal R$ becomes large, whereas increasing $C_\Sigma$ shifts the optimal frequency downward.

\subsection{Device Fabrication Process}
\label{sec:fabrication}
\begin{figure}[!htbp]
    \centering
    \includegraphics[width=1\linewidth]{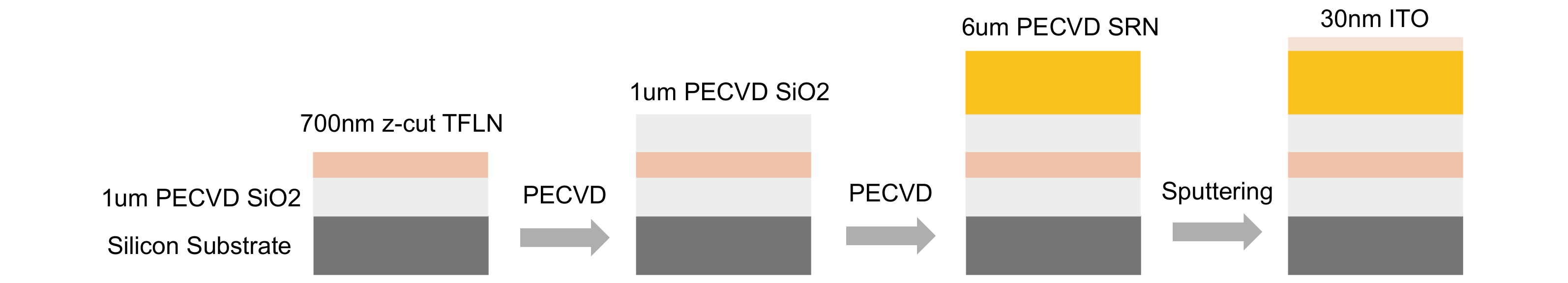}
    \caption{\textbf{Schematics of the fabrication process for our device.}}
    \label{fig:S3}
\end{figure}

The fabrication process is summarized in Fig.~\ref{fig:S3}. We fabricated the device starting from a commercial thin-film LN wafer (NanoLN). The wafer consists of a p-type silicon substrate with a resistivity of \SIrange{0.01}{0.05}{\ohm\centi\metre}, coated with a \SI{1}{\micro\metre} silicon dioxide layer deposited by PECVD, and a \SI{700}{\nano\metre} z-cut MgO-doped lithium niobate film bonded using the ion-cut process. The wafer was first diced into \SI{1}{\centi\metre} $\times$ \SI{1}{\centi\metre} chips with a Disco dicing saw for subsequent device processing. A further \SI{1}{\micro\metre} silicon dioxide layer was then deposited by PECVD to serve as the cladding, followed by deposition of a \SI{6}{\micro\metre} SRN layer, which functions as the photoconductive material. The SRN film was deposited in an Oxford Plasmalab~100 system using gas flows of \SI{8}{\sccm} SiH$_4$, \SI{40}{\sccm} H$_2$, and \SI{2000}{\sccm} N$_2$ at \SI{350}{\celsius} and \SI{1900}{\milli\torr}.

Indium tin oxide (ITO) electrodes were deposited on the chip using a PVD75 sputtering system. The chip edges were masked with tape during deposition, in order to suppress dielectric breakdown through air between the top electrode and the conductive Si substrate. Taping the chip edges leaves a \SI{1}{\milli\metre}-wide ITO-free margin at the edges, which lengthens the air path between the electrode and the substrate. To reduce optical coupling loss at the facets, the chip facets were polished with an Allied MultiPrep polisher using diamond lapping films of successively finer grit (\SI{6}{\micro\metre}, \SI{1}{\micro\metre}, and \SI{0.5}{\micro\metre}).

\section{Device Performance}

\subsection{Programmable Index Change Measurement}
\label{sec:holography}
As described in Section~1B of the Supplementary Information of Ref.~\cite{Onodera2025}, we used an off-axis holography method to characterize the refractive-index modulation of the device as a function of the applied electrical bias and the illumination intensity on the photoconductive layer. A probe beam was first sent through a polarizer to align its polarization with the $y$ axis of the waveguide. It was then coupled into the waveguide through a microscope objective to excite the $\mathrm{TM_{0}}$ mode. The output facet was imaged through an objective onto a near-infrared (NIR) camera operating at \SI{4}{\kilo\hertz}. To retrieve the phase accumulated through the device, the output beam was interfered with an off-axis reference beam split from the same input source by means of a beam splitter, generating spatial interference fringes suitable for holographic phase reconstruction. All measurements reported here are referenced to the effective index of the same mode in the absence of an externally applied voltage. For frequencies above \SI{2}{\hertz}, we used stroboscopic sampling to acquire more points per cycle. We set the frame rate of the camera to be a multiple of the drive frequency plus \SI{0.1}{\hertz}, and reconstructed the phase change pattern over a full cycle.

\begin{figure}[!htbp]
    \centering
    \includegraphics[width=1\linewidth]{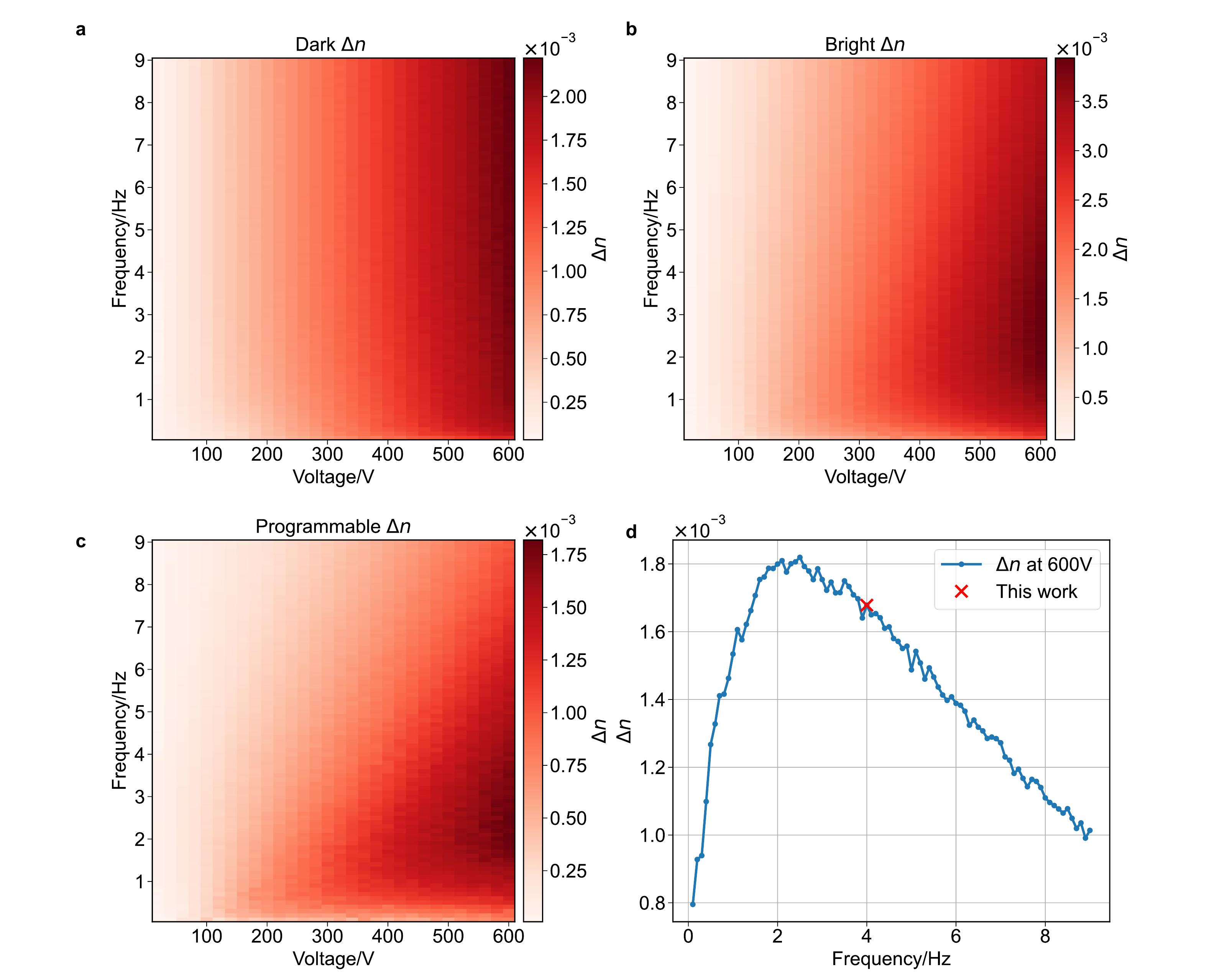}
    \caption{\textbf{Measured index change versus applied voltage and drive frequency.} \textbf{a}, AC index-shift amplitude as a function of AC drive voltage and frequency in the dark state. \textbf{b}, AC index-shift amplitude as a function of AC drive voltage and frequency in the bright state. \textbf{c}, Programmable index change between the bright and dark states. \textbf{d}, Programmable index change at an applied voltage of \SI{600}{\volt} as a function of drive frequency. The red cross marks the programmable index change achieved in this work.}
    \label{fig:Delta n scan}
\end{figure}

\par Unlike in our previous work, here we characterized the index modulation as a function of the drive frequency, which allowed us to choose an operating frequency that balances index contrast against speed. The results are shown in Fig.~\ref{fig:Delta n scan}. We observed that the refractive-index change in the dark state approaches saturation at high frequencies, while it decreases at high frequencies in the bright state. As a result, we observed an optimal drive frequency that maximizes the programmable contrast, as predicted in Appendix~\ref{sec:design}. We chose a driving frequency of \SI{4}{\hertz} where we measured a programmable index modulation of approximately $1.7\times 10^{-3}$ at an applied voltage of \SI{600}{\volt}.

\subsection{Insertion Loss and Crosstalk Measurements}
\label{sec:insertion}

We report two insertion-loss figures.
First, we report the total free-space-input to free-space-output loss of the system. This quantity includes multiple contributions: coupling losses at the input and output facets due to reflection, scattering, and mode-mismatch, and absorption and radiation losses inside the device. 
To measure the total loss, we operated the device without illumination and launched an input beam with a waist of approximately \SI{300}{\micro\metre} and a power of \SI{2.1}{\micro\watt}. We measured an output power of \SI{620}{\nano\watt}, corresponding to a total insertion loss of \SI{5.3}{\decibel}.

Second, we estimate the intrinsic insertion loss of the programmable region, i.e., the loss seen by light already guided in an index-matched TFLN waveguide. These values are reported in the main text and figures, and they are meant to be comparable to values usually reported in the inverse-design literature, where losses due to free-space-to-chip or fiber-to-chip interfaces are commonly subtracted from the final reported insertion loss; for example, Ref.~\cite{sunEdgeguidedInverseDesign2025} reports \SIrange{5}{6}{\decibel} per facet but a much lower device insertion loss. 

A direct measurement of this ``intrinsic'' insertion loss is not possible in the present free-space experimental geometry, because the loss of the programmable region cannot be separated from other losses in the optical path, such as scattering at the facets. We also do not image the phase of the output light or measure the output power (except for the \SI{5.3}{\decibel} measurement reported above, which was performed with a photodiode). Instead, we rely on the following approximation to estimate the insertion loss through the programmable waveguide area:

\newcommand{\Eout}[1][]{E_{\mathrm{out}#1}}
\newcommand{\Ein}[1][]{E_{\mathrm{in}#1}}
\newcommand{\tEout}[1][]{\widetilde{E}_{\mathrm{out}#1}}
\newcommand{\Etgt}[1][]{E_{\mathrm{t}#1}}
\newcommand{\IL}{\mathit{IL}}
\newcommand{\tIL}{\widetilde{\IL}}
\newcommand{\CT}{\mathit{CT}}
\newcommand{\tCT}{\widetilde{\CT}}
\newcommand{\Pmax}{P_{\max}}
\newcommand{\Ptot}[1][]{P^{\mathrm{tot}}_{\mathrm{out}#1}}
\begin{align}
  \tIL_{i} = -10\log_{10}
    \frac{\bigl|\ip{\Etgt[,i]}{\tEout[,i]}\bigr|^{2}}
         {\Pmax\,\ip{\Etgt[,i]}{\Etgt[,i]}},
\end{align}

where $\Pmax=\max_{i}\ip{\tEout[,i]}{\tEout[,i]}$, the maximum is taken across all input \unitforN{}s $i$, and we use the notation $\ip{A}{B}=\int A^*(x)B(x)\,\mathrm{d}x$ for the inner product. The tilde indicates that $\tIL$ is an estimate of the ``real'' insertion loss, as defined below. $\tEout[,i]$ refers to the dimensionless output field estimated from the intensity measurement on the camera: $\tEout[,i] := \sqrt{I_{\mathrm{out},i}}$. This definition allows us to estimate the insertion loss from dimensionless intensity measurements on our camera. We compare this to a hypothetical programmable waveguide that is fully integrated with mode-matched photodetectors that measure a perfectly accurate insertion loss:

\begin{align}
  \IL_{i} = -10\log_{10}\frac{P_\mathrm{out,i}}{P_\mathrm{in,i}} = -10\log_{10}
    \frac{\bigl|\ip{\Etgt[,i]}{\Eout[,i]}\bigr|^{2}}
         {\bigl|\ip{\Ein[,i]}{\Ein[,i]}\bigr|^{2}},
\end{align}

where the target power is equal to the input power.
Rewriting our insertion loss estimate, we find that:

\begin{align}
    \tIL_{i} &= -10\log_{10} \Bigg[
    \underbrace{\frac{\bigl|\ip{\Etgt[,i]}{\Eout[,i]}\bigr|^{2}} {\bigl|\ip{\Ein[,i]}{\Ein[,i]}\bigr|^{2}}}_{\mathrm{accurate}}
    \underbrace{\frac{\bigl|\ip{\Etgt[,i]}{\tEout[,i]}\bigr|^{2}} {\bigl|\ip{\Etgt[,i]}{\Eout[,i]}\bigr|^{2}}}_{\mathrm{phase-mismatch}}
    \underbrace{\frac{\bigl|\ip{\Ein[,i]}{\Ein[,i]}\bigr|^{2}} {\ip{\Etgt[,i]}{\Etgt[,i]}\Pmax}}_{\mathrm{absorption, radiation}} \Bigg] \\
    &= \IL_{i} - \IL_{\textrm{phase-mismatch}} - \IL_{\textrm{absorption, radiation}},
\end{align}

where the loss is decomposed into separate terms, each of which is non-negative. We see that the insertion loss reported in the main text, $\tIL_i$, is a lower bound on the real insertion loss $\IL_{i}$. We estimate that the absorption and radiation term is less than \SI{1}{\decibel} for our \SI{1}{\centi\meter}-long device~\cite{Onodera2025}. The phase-mismatch term could be accurately measured with phase-imaging methods in future work. 

We use a similar formula to estimate the crosstalk between input \unitforN{} $i$ and undesired output \unitforN{} $j$ (the target of input $j$):
\begin{align}
  \tCT_{ij} = 10\log_{10} \frac{\bigl|\ip{\Etgt[,j]}{\tEout[,i]}\bigr|^{2}}{\Pmax\,\ip{\Etgt[,j]}{\Etgt[,j]}} + \tIL_{j}.
\end{align}
This should be compared with the conventional definition:
\begin{equation}
  \CT_{ij} = 10\log_{10}
    \frac{P_{j\leftarrow i}}{P_{j\leftarrow j}},
  \qquad
  P_{j\leftarrow i}
    = \frac{\bigl|\ip{\Etgt[,j]}{\Eout[,i]}\bigr|^{2}}{\ip{\Etgt[,j]}{\Etgt[,j]}}.
  \label{eq:ctstrict}
\end{equation}
A calculation analogous to the one above, comparing $\tCT_{ij}$ with $\CT_{ij}$, shows that our estimates for crosstalk values are accurate up to errors induced by the assumption of a flat phase-front for $\Eout$. However, since both $P_{j\leftarrow i}$ (in the numerator) and $P_{j\leftarrow j}$ (in the denominator) will carry errors, the difference between $\tCT_{ij}$ and $\CT_{ij}$ can be positive or negative, and the crosstalk values we report cannot easily be interpreted as upper or lower bounds on the real crosstalk values.
\subsection{Device Endurance}
\label{sec:Endurance}

\begin{figure}[!htbp]
    \centering
    \includegraphics[width=1\linewidth]{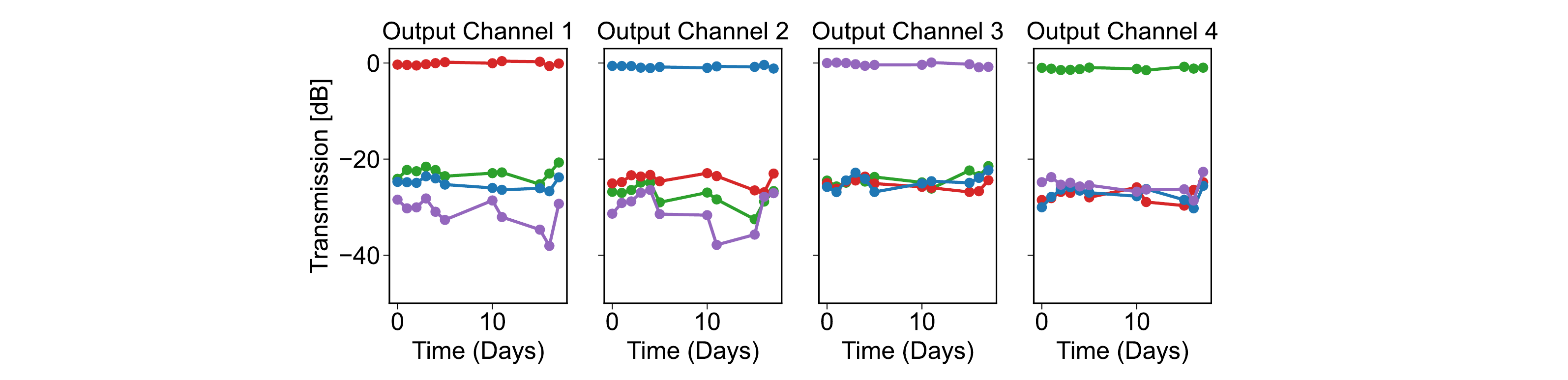}
    \caption{\textbf{Trained $4\times4$ permutation device performance as a function of the time after training.} The device shows stable performance over a time span of two weeks.}
    \label{fig:Endurance test}
\end{figure}

We measured the endurance of the designed photonic functions by repeatedly applying the same trained illumination pattern over a time window of approximately two weeks~(Fig.~\ref{fig:Endurance test}). For each measurement, we applied the trained pattern onto the device and recorded the corresponding output. We returned the device to a uniform illumination immediately afterwards. The performance remained nearly unchanged over the time window, showing that the device response to the inverse-designed illumination pattern is stable and persistent within this timescale.

However, we note that the device's output intensity profile will be distorted if kept under a fixed non-uniform illumination pattern for a long time (e.g., 24 hours). We observed a residual index profile of amplitude ${\sim}10^{-4}$ inside the device. We refer to this as the ``burned-in'' index profile. This residual index pattern distorts the output intensity profile and therefore the previously trained digital model no longer accurately represents the device.

We observed this ``burned-in'' effect by examining the output intensity profile of the device under constant illumination, which was strongly distorted after prolonged illumination. The residual index pattern was later measured by launching a large number of random input wavefronts, recording the corresponding output intensity distributions, and reconstructing the index profile from these data.

\par We performed a series of controlled experiments to identify the origin of this effect. Starting from a clean device, i.e., a device that performs as expected, we applied a grating illumination pattern with external voltage on (the standard experimental configuration) for around 24 hours, and we later observed a persistent residual grating index profile in the device. However, when we applied the same constant illumination pattern for a long time, but with voltage off, we observed no such residual index profile. This indicates that the ``burned-in'' effect requires both constant illumination and external voltage bias at the same time.

\par When we started from a ``burned-in'' state, i.e. the residual index profile had been created in the device, we applied uniform illumination with the external voltage on for 24 hours, and we found that the residual pattern was removed and a clean state was restored. For comparison, we started from a ``burned-in'' state again, but we left the device in the dark with the external voltage on for a similar duration, and we did not observe a change to the ``burned-in'' pattern. With both voltage and illumination off, the residual profile was not observable and the output intensity of the device returned to normal, but it reappeared when the bias was restored. This shows that the ``burned-in'' index profile is not a static index offset but appears only with the external voltage on. 

\par Although the exact reason for this effect is unclear, we attribute it to persistent photoconductivity. which we have observed via electrical measurements on earlier samples under UV illumination.
If a device is exposed to a constant nonuniform illumination with external voltage, it could produce a correspondingly nonuniform persistent photoconductive state in the photoconductive layer. This state changes the local voltage distribution across the lithium niobate layer when an external voltage is applied, therefore producing a residual electro-optic index pattern through the Pockels effect. Removing the voltage also removes the electro-optic index change, so the ``burned-in'' pattern is no longer observable in the output intensity, but the underlying photoconductive state remains and the residual index profile can be observed again by reapplying the external voltage. Uniform illumination under external voltage would erase the pattern by driving the photoconductive or trapped-charge distribution toward a spatially uniform state. 

\subsection{Summary of demonstrated and projected performance}
\label{sec:performance}
Table~\ref{tab:performance_comparison} summarizes the demonstrated performance and the projected performance of improved devices. Projections for improved versions of the present stack are discussed in Appendices~\ref{sec: Future Device} and~\ref{sec: Polymer Device}, for DC devices in Appendix~\ref{sec:DC device}, and for non-volatile devices in Appendix~\ref{sec: Non-volatile}.

\begin{table}[H]
\centering
\setlength{\tabcolsep}{8pt}
\renewcommand{\arraystretch}{1.2}
\begin{tabular}{|l|c|c|c|c|}
\hline
\textbf{Performance} & \textbf{This work} & \pbox{5cm}{\textbf{Improved} \\ \textbf{current stack}} & \pbox{5cm}{\textbf{Future} \\ \textbf{DC device}} & \pbox{5cm}{\textbf{Future} \\ \textbf{non-volatile device}}\\
\hline
\shortstack[l]{\rule{0pt}{2ex}Programmable index \\modulation $\Delta n$}
& \raisebox{1.0ex}{$1.7\times 10^{-3}$~$^{\mathrm{a}}$}
& \raisebox{1.0ex}{$1.5\times10^{-2}$~$^{\mathrm{b}}$}
& \raisebox{1.0ex}{$7\times 10^{-2}$~$^{\mathrm{c}}$}
& \raisebox{1.0ex}{$10^{-2}$~$^{\mathrm{d}}$} \\
\hline
\shortstack[l]{Input--output\\recording speed}
& \SI{4}{\hertz}$~^{\mathrm{e}}$
& Not device-limited~$^{\mathrm{f}}$
& Not device-limited~$^{\mathrm{g}}$
& Not device-limited~$^{\mathrm{h}}$\\
\hline
\shortstack[l]{\rule{0pt}{2ex}Pattern \\update time}
& \raisebox{1.0ex}{\SI{1.5}{\second}~$^{\mathrm{i}}$}
& \raisebox{1.0ex}{\SI{1}{\milli\second}~$^{\mathrm{j}}$}
& \raisebox{1.0ex}{\SI{1}{\second}~$^{\mathrm{k}}$}
& \raisebox{1.0ex}{\SI{1}{\second}~$^{\mathrm{l}}$}\\
\hline
Illumination resolution
& \SI{9}{\micro\metre}~$^{\mathrm{m}}$
& \SI{3}{\micro\metre}~$^{\mathrm{n}}$
& \SI{1}{\micro\metre}~$^{\mathrm{o}}$
& \SI{1}{\micro\metre}~$^{\mathrm{p}}$ \\
\hline
\shortstack[l]{\rule{0pt}{2ex}Area of programmable \\index $(z \times x)$}
& \raisebox{1.0ex}{\SI{9}{\milli\metre} $\times$ \SI{8}{\milli\metre}~$^{\mathrm{q}}$}
& \raisebox{1.0ex}{\SI{5}{\centi\metre} $\times$ \SI{5}{\centi\metre}~$^{\mathrm{r}}$}
& \raisebox{1.0ex}{\SI{5}{\centi\metre} $\times$ \SI{5}{\centi\metre}~$^{\mathrm{s}}$}
& \raisebox{1.0ex}{\SI{5}{\centi\metre} $\times$ \SI{5}{\centi\metre}~$^{\mathrm{t}}$}\\
\hline
\shortstack[l]{\rule{0pt}{2ex}Illumination power \\ to maintain pattern}
& \raisebox{1.0ex}{\SI{3}{\milli\watt}~$^{\mathrm{u}}$}
& \raisebox{1.0ex}{\SI{1}{\micro\watt}~$^{\mathrm{v}}$}
& \raisebox{1.0ex}{\SI{3}{\milli\watt}~$^{\mathrm{w}}$}
& \raisebox{1.0ex}{\SI{0}{\milli\watt}~$^{\mathrm{x}}$} \\
\hline
Electrical power consumption
& \SI{2}{\milli\watt}~$^{\mathrm{y}}$
& \SI{1}{\micro\watt}~$^{\mathrm{z}}$
& \SI{2}{\milli\watt}~$^{\mathrm{a'}}$
& \SI{2}{\milli\watt}~$^{\mathrm{b'}}$\\
\hline
\end{tabular}
\caption{
\textbf{Performance comparison between this work and potential improved future devices.} Columns list, from left to right, the performance metric, the value achieved in this work, the projected value for a future improved version of the present device, the projected value for a DC-driven device, and the projected value for a non-volatile ferroelectric device. \textbf{a}, See Appendix~\ref{sec:holography}. \textbf{b}, Projected for the improved stack; see Appendices~\ref{sec: Future Device} and~\ref{sec: Polymer Device}. \textbf{c}, See Appendices~\ref{sec:DC device} and~\ref{sec: Polymer Device}. \textbf{d}, See Appendix~\ref{sec: Non-volatile}. \textbf{e}, Chosen to balance index contrast against speed; see Appendix~\ref{sec:holography}. \textbf{f--h}, Input--output samples can be collected within a phase window over which the programmed transformation is approximately constant. Many samples can be processed during a single AC cycle. The recording speed is therefore not fundamentally limited by the AC frequency, but is instead determined primarily by the bandwidths of the external input modulator and output detector. \textbf{i}, The response time was measured by switching patterns at increasing rates and determining the fastest rate at which the output remained stable. \textbf{j}, See Appendix~\ref{sec: Future Device}. \textbf{k}, See Appendix~\ref{sec:DC device}. \textbf{l}, See Appendix~\ref{sec: Non-volatile}. \textbf{m}, Determined from the \SI{13.7}{\micro\metre} pixel pitch of our DMD (DLP Subsystem V-7000 VIS) and the imaging magnification. \textbf{n}, Assumes an improved photoconductor and finer projection optics; see Appendix~\ref{sec: Future Device}. \textbf{o}, See Appendix~\ref{sec:DC device}. \textbf{p}, See Appendix~\ref{sec: Non-volatile}. \textbf{q}, The sizes are measured using the imaged pixel size of DMD pixels and camera pixels in the illumination setup. \textbf{r}, The estimate assumes a larger DMD resolution ($7680 \times 4320$) and combining multiple DMDs in the illumination setup; the ultimate limit is then the wafer size. \textbf{s}, Same as the improved device. \textbf{t}, Same as the improved device. \textbf{u}, The power was measured using a power meter on top of our device. The illumination power of a fully on DMD was measured to be \SI{50}{\milli\watt}, of which 13\% of the pixels were used, at an average of 50\% of maximum intensity. \textbf{v}, See Appendix~\ref{sec: Future Device}. \textbf{w}, See Appendix~\ref{sec:DC device}. \textbf{x}, See Appendix~\ref{sec: Non-volatile}. \textbf{y}, The average real power was calculated from the measured peak voltage $V_{\mathrm{pk}}=\SI{600}{\volt}$, peak current $I_{\mathrm{pk}}=\SI{25}{\micro\ampere}$, and voltage--current phase difference $\phi\approx0.4\pi$ using $P_{\mathrm{HV}}=\tfrac{1}{2}V_{\mathrm{pk}}I_{\mathrm{pk}}\cos\phi\approx\SI{2.3}{\milli\watt}$, reported here as approximately \SI{2}{\milli\watt}. \textbf{z}, See Appendix~\ref{sec: Future Device}. \textbf{a$'$}, See Appendix~\ref{sec:DC device}. \textbf{b$'$}, See Appendix~\ref{sec: Non-volatile}.}
\label{tab:performance_comparison}
\end{table}

\begin{table}[H]
\centering
\setlength{\tabcolsep}{4pt}
\renewcommand{\arraystretch}{1.2}
\begin{tabular}{|l|c|c|c|c|c|c|}
\hline

\raisebox{1.3ex}{\textbf{\rule{0pt}{3.5ex}Device Type}}
& \raisebox{1.3ex}{\textbf{Platform}}
& \textbf{\shortstack[c]{Device\\Scale}}
& \raisebox{1.3ex}{\textbf{Crosstalk}}
& \textbf{\shortstack[c]{Insertion\\Loss}}
& \textbf{\shortstack[c]{Bandwidth\\/$\lambda$ Spacing$^{a}$}}
& \raisebox{1.3ex}{\textbf{Reference}} \\

\hline
\multirow{6}{*}{\shortstack[l]{Permutations\\/Routers}}
& Programmable device
& $4\times4$
& \SI{-27}{\decibel}
& \SI[parse-numbers=false]{(1.4 + 5.3)}{\decibel}$^{b}$
& \SI{100}{\nano\metre}$^{c}$
& This work \\
\cline{2-7}
& Programmable device
& $8\times8$
& \SI{-13}{\decibel}
& \SI[parse-numbers=false]{(4 + 5.3)}{\decibel}$^{b}$
& \SI{100}{\nano\metre}$^{c}$
& This work \\
\cline{2-7}
& EBL SOI Waveguide Crossing
& $4\times 4$
& \SI{-23}{\decibel}
& \SI{0.75}{\decibel}
& \SI{40}{\nano\metre}
& Ref.~\cite{luInversedesignedUltracompactStarcrossings2017} \\
\cline{2-7}
& EBL SOI Waveguide Crossing
& $6\times 6$
& \SI{-18}{\decibel}
& \SI{1.5}{\decibel}
& \SI{40}{\nano\metre}
& Ref.~\cite{luInversedesignedUltracompactStarcrossings2017} \\
\cline{2-7}
& EBL SOI Waveguide Crossing
& $2\times 2$
& \SI{-35}{\decibel}
& \SI{0.3}{\decibel}
& \SI{60}{\nano\metre}
& Ref.~\cite{xu5x5Mm2Compact2017} \\
\cline{2-7}
& EBL LNOI Waveguide Crossing
& $2\times 2$
& \SI{-36}{\decibel}
& \SI{0.48}{\decibel}
& \SI{100}{\nano\metre}
& Ref.~\cite{shangInverseDesignedLithiumNiobate2023} \\
\noalign{\hrule height 1pt}

\multirow{6}{*}{\shortstack[l]{Mode \\Demultiplexer}}
& Programmable device
& $4\times4$
& \SI{-13}{\decibel}
& \SI[parse-numbers=false]{(2.2 + 5.3)}{\decibel}$^{b}$
& \SI{100}{\nano\metre}$^{c}$
& This work \\
\cline{2-7}
& EBL SOI Y-junction
& $3\times 3$
& \SI{-19}{\decibel}
& \SI{2.5}{\decibel}
& \SI{80}{\nano\metre}
& Ref.~\cite{changUltracompactModeMultiplexer2018} \\
\cline{2-7}
& EBL SOI Waveguide
& $4\times 4$
& \SI{-20}{\decibel}
& \SI{1.5}{\decibel}
& \SI{60}{\nano\metre}
& Ref.~\cite{xieHighlyCompactEfficient2020} \\
\cline{2-7}
& EBL SOI Waveguide
& $4\times 4$
& \SI{-18.89}{\decibel}
& \SI{0.8}{\decibel}
& \SI{120}{\nano\metre}
& Ref.~\cite{yangMultidimensionalDataTransmission2022} \\
\cline{2-7}
& EBL SOI Metamaterial Waveguide
& $5\times 5$
& \SI{-20.72}{\decibel}
& \SI{1.97}{\decibel}
& \SI{35}{\nano\metre}
& Ref.~\cite{sunEdgeguidedInverseDesign2025} \\
\cline{2-7}
& EBL SOI Gratings
& $6\times 6$
& \SI{-18}{\decibel}
& \SI{0.77}{\decibel}
& \SI{100}{\nano\metre}
& Ref.~\cite{wuCompactScalableMode2025} \\
\noalign{\hrule height 1pt}

\multirow{5}{*}{\shortstack[l]{Wavelength \\Demultiplexer}}
& Programmable device
& $2\times2$
& \SI{-8.4}{\decibel}
& \SI[parse-numbers=false]{(1.7 + 5.3)}{\decibel}$^{b}$
& \SI{240}{\nano\metre}
& This work \\
\cline{2-7}
& EBL SOI Waveguide
& $2\times 2$
& \SI{-11}{\decibel}
& \SI{1.8}{\decibel}
& \SI{250}{\nano\metre}
& Ref.~\cite{piggottInverseDesignDemonstration2015} \\
\cline{2-7}
& EBL SiN Waveguide
& $2\times 2$
& \SI{-3}{\decibel}
& \SI{1.9}{\decibel}
& \SI{253}{\nano\metre}
& Ref.~\cite{biInversedesignedSiliconNitride2026} \\
\cline{2-7}
& EBL SOI Waveguide
& $3\times 3$
& \SI{-10.7}{\decibel}
& \SI{2.29}{\decibel}
& \SI{40}{\nano\metre}
& Ref.~\cite{suInverseDesignDemonstration2018} \\
\cline{2-7}
& EBL SOI Waveguide
& $4\times 4$
& \SI{-19}{\decibel}
& \SI{2}{\decibel}
& \SI{20}{\nano\metre}
& Ref.~\cite{cheungInversedesignedCWDMDemultiplexer2024} \\
\hline
\end{tabular}
\caption{\textbf{Performance comparison between this work and other works using traditional inverse-design platforms.} Columns list device category, device/technology, device scale (input \unitforN{}s $\times$ output \unitforN{}s), crosstalk, insertion loss, bandwidth or wavelength spacing, and reference. Crosstalk and insertion loss are reported for \SI{1550}{\nano\metre} (except for the wavelength demultiplexer task). EBL: electron-beam lithography. SOI: silicon-on-insulator. LNOI: lithium-niobate-on-insulator. SiN: silicon nitride. \textbf{a}, $\lambda$ spacing here refers specifically to wavelength-demultiplexer devices, which are designed to separate light of different wavelengths. \textbf{b}, As described in Appendix~\ref{sec:insertion}, we report \SI{5.3}{\decibel} as an estimate for absorption, radiation, and facet-coupling losses from free-space inputs to free-space outputs. The other term is a lower-bound estimate of coupling losses due to mode mismatch. \textbf{c}, Our bandwidth measurement is limited by the tuning range of our tunable laser; our simulation results show that the bandwidth could extend to ${\sim}\SI{800}{\nano\metre}$ (Appendix~\ref{sec:broadband}).}
\label{tab:performance_comparison_lit}
\end{table}

\section{Experimental Setup}
The experimental setup we used was largely identical to that of our previous work; details are provided in Section~2 of the Supplementary Information of Ref.~\cite{Onodera2025}. We used a spatial light modulator (SLM, Meadowlark Optics UHSP1K-850-1650-PC8)-based 1D spatial beam shaper for generating arbitrary optical input fields, and a projector based on a digital micromirror device (DMD, Vialux V-7000) illuminated by a high-power LED (Thorlabs SOLIS-525C) to project illumination patterns onto the surface of the 2D-programmable waveguide. In the beam shaper, we created spatially varying phase gratings across the two-dimensional phase panel of the SLM to encode arbitrary one-dimensional amplitude and phase distributions on the input facet. The coherent laser light from a Santec TSL-570 (for C-band light) or a distributed-feedback fiber-pigtailed laser diode module (DFL Series, SS63310100007, Lasertel) (for \SI{1310}{\nano\metre}) formed diffraction orders with the first order carrying the designed input optical field at the focal plane after the first lens (Thorlabs A1380-C-ML). The light was spatially filtered to remove the zeroth order and demagnified by a $4f$ system with a lens (Thorlabs TTL200-A, $f=\SI{200}{\milli\metre}$) and an objective (Olympus UPLFLN 40X Objective, $\mathrm{NA}=0.75$, $f=\SI{4.5}{\milli\metre}$). The first order was then coupled into the input facet of the device. In parallel, we used the DMD and the green LED to project programmable illumination patterns to the waveguide top surface through a second $4f$ imaging system (using two lenses, Thorlabs TL300-A and Thorlabs TTL200-A), with a Dove prism (Thorlabs PS993M) introduced to compensate for the $45^\circ$ rotation of the micromirror array. The output intensity from the chip was collected using an infrared camera (Allied Vision Goldeye G-033 SWIR TEC1). 

\label{sec:setup}

\section{Digital Model and Inverse Design Process}

\subsection{Hybrid Data-Driven Digital Model}
\begin{figure}[!htbp]
    \centering
    \includegraphics[width=1\linewidth]{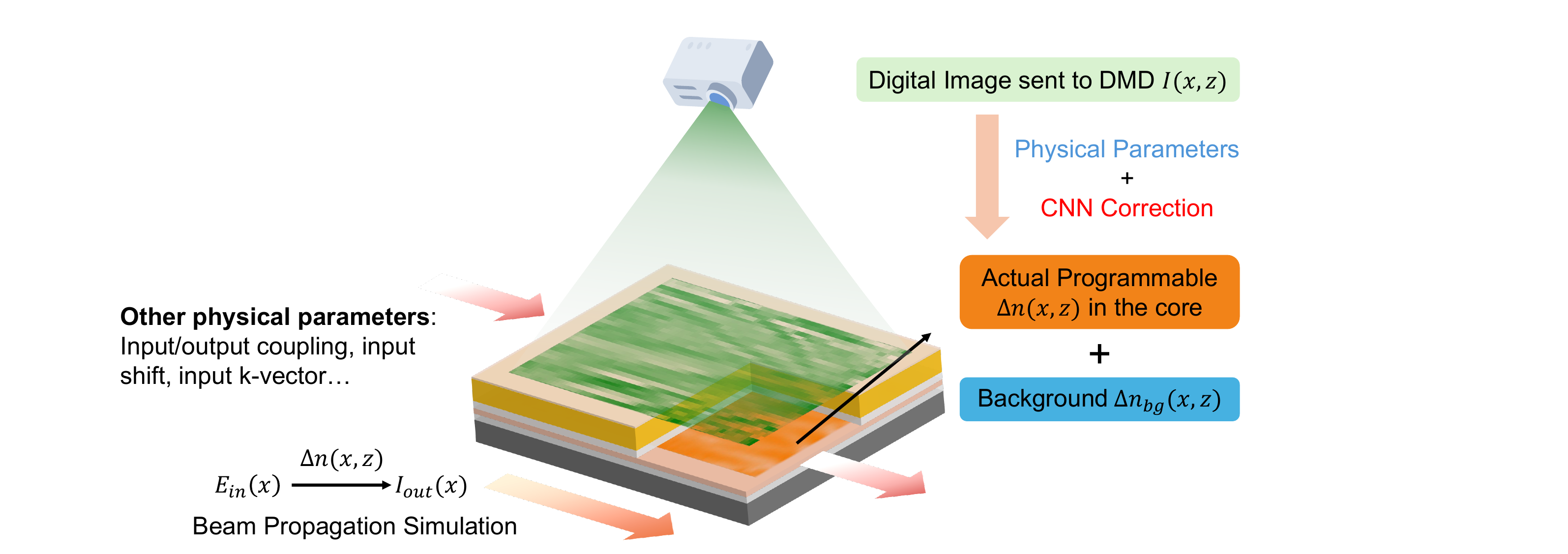}
    \caption{\textbf{Schematic of the structure of our digital model.} The model is a 2D beam-propagation simulation with additional physical parameters: input/output coupling coefficient, input overall spatial shift, and overall $k$-vector. We also use a background index pattern to account for static wavefront distortion in the LN core. For the illumination and programmable index pattern, we use a Gaussian-blurred mapping as a physically parametrized model, while adding a correction layer based on a convolutional neural network (CNN) to compensate for the simulation--reality gap.}
    \label{fig:DT illustration}
\end{figure}

In this section, we describe the digital model we created for the 2D programmable photonic device. The model combines a physics-based simulation, parameterized by measured device properties, with a data-driven neural-network correction. A simplified illustration of the model structure is shown in Fig.~\ref{fig:DT illustration}.

First, we introduce the theoretical framework for simulating the two-dimensional programmable waveguide. For the TM-polarized case considered here, the relevant electric-field component is oriented along the $y$ direction, and the field $\tilde{E}$ therefore satisfies
\begin{equation}
\frac{\partial^2 \tilde{E}}{\partial z^2} + \frac{\partial^2 \tilde{E}}{\partial x^2} + n^2(x,z)k_0^2 \tilde{E} = 0,
\label{eq:helmholtz_2d}
\end{equation}
where $n(x,z)$ denotes the effective refractive-index distribution of the slab and $k_0$ is the free-space wavenumber. This simplified two-dimensional description is appropriate for our system because the induced refractive-index changes are small and there is only one confined propagation mode in the stack. For numerical efficiency, we write the field as $\tilde{E}(x,z)=E(x,z)\exp(ikz)$ and apply the slowly varying envelope approximation, which yields
\begin{equation}
\frac{\partial E}{\partial z}
=
\frac{i}{2k}\frac{\partial^2 E}{\partial x^2}
+i\,\Delta n(x,z)\,k_0 E,
\label{eq:bpm_main}
\end{equation}
where $\Delta n(x,z)=n(x,z)-n_0$ is the refractive-index perturbation relative to a reference index $n_0$, with $\Delta n \ll n_0$, and $k=n_0k_0$ is the wavenumber in the waveguide. Equation~\eqref{eq:bpm_main} neglects back-reflection, which is justified because $\Delta n \ll n_0$ and the targeted functions are non-resonant. We solve this evolution equation numerically using the split-step Fourier method. As in our previous work, we model the mapping from the illumination intensity distribution $I(x,z)$ to the corresponding refractive-index change inside the lithium niobate core, $\Delta n(x,z)$:
\begin{equation}
    \Delta n(x,z) = F(I)(g(x,z)\ast I(x,z)),
\end{equation}
where $F(I)$ represents the nonlinear mapping from illumination to index change, $\ast$ denotes two-dimensional convolution, and $g$ is a Gaussian kernel. Even after fine-tuning, the purely physics-based model did not generalize well across tasks. We attribute this to inaccurate modeling of the illumination-to-index mapping. To achieve a more accurate fit of the conversion, we introduced a convolutional neural network (CNN) to correct the gap between the physics-based and actual index profiles, making the model a hybrid digital model.

\label{sec:DT illustration}

\par Although the experimental setup underwent a series of calibrations, as described in our previous work, our model still contains other physical parameters, including input shift in the $x$ direction, overall input angle, input coupling efficiency and phase modulation, point-spread function adjusted for the free-space beam-shaper setup, and output coupling efficiency.

\subsection{Generalized Dataset}
\label{sec:dataset}
\par
\begin{figure}[!htbp]
    \centering
    \includegraphics[width=1\linewidth]{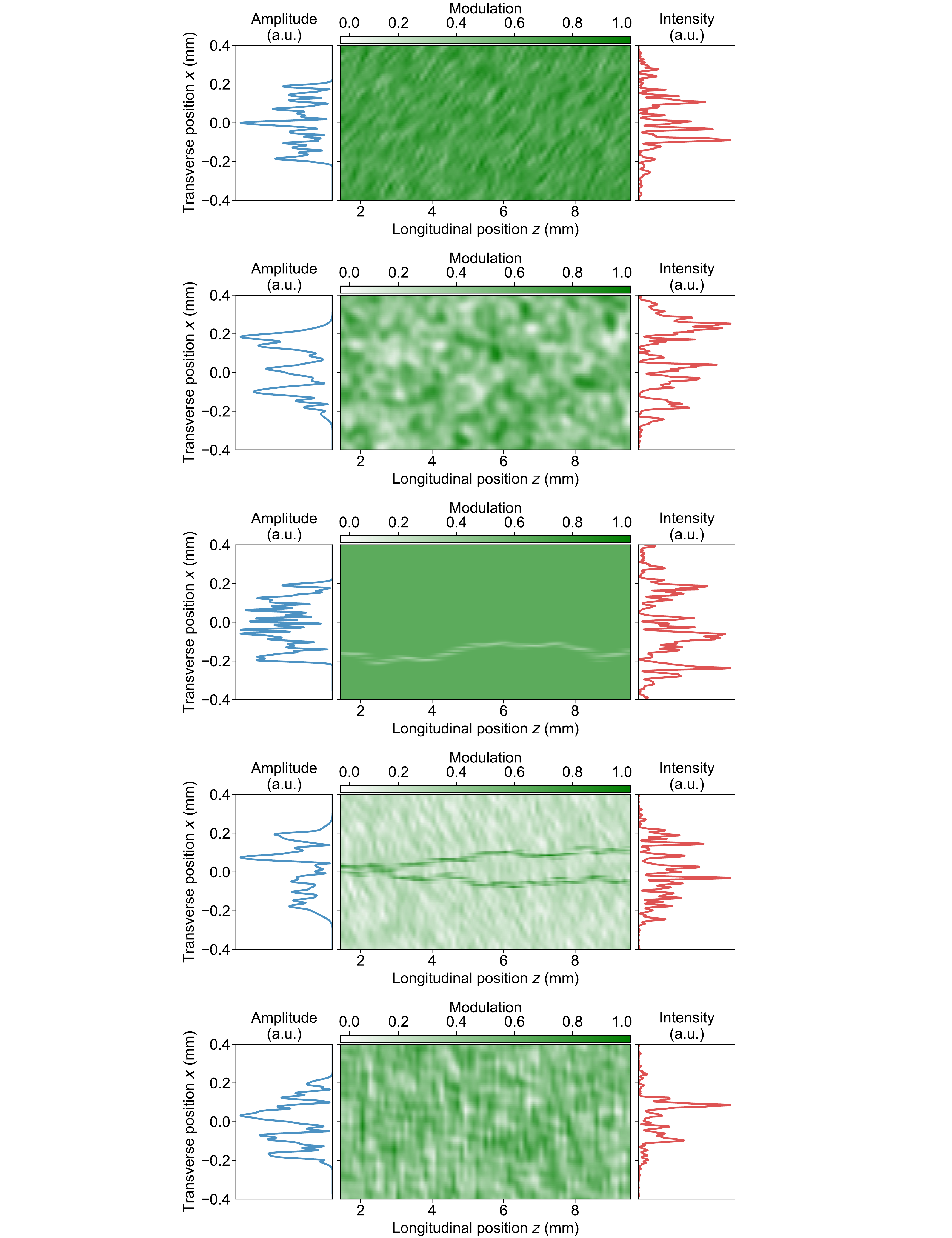}
    \caption{\textbf{Five random samples from the dataset used for tuning the digital model.}}
    \label{fig:Dataset_sample}
\end{figure}

 We built a generalized input--output dataset that represents our device behavior for the purpose of inverse-designing various distinct photonic functions. Our dataset has two categories of input parameters: 1D input field and 2D illumination patterns. Representative samples are shown in Fig.~\ref{fig:Dataset_sample}.

\par We generated each random 1D input field as a superposition of several random complex components. The number of components was drawn randomly for each sample. Each component was then initialized as complex white noise confined to a finite window in $x$. To control the smallest feature size and corresponding $k$-vector distribution, the noise was Fourier transformed and filtered by a Gaussian low-pass filter in spatial-frequency space, with bandwidth $\Delta f = 1/L_f$, where the characteristic feature size $L_f$ was also randomly sampled. After inverse Fourier transformation, we normalized each component and assigned it a random amplitude. The components were then summed coherently. Finally, a $\pi$ phase step was inserted at a random transverse position by multiplying one side of the field by $-1$. The final field was then normalized by its peak magnitude. This procedure produces spatially smooth but highly diverse complex input beams with randomized local structure, relative phase, and occasional phase discontinuities.

\par For the illumination patterns, we generated random two-dimensional illumination patterns by superposing a small number of independently sampled layers in the $(x, z)$ plane. Each pattern consisted of one to four layers (number drawn uniformly). We assigned each layer to one of three classes: blob-like patterns, streak-like patterns, or waveguide-like patterns. Blob and streak layers were created in the spatial-frequency domain by filtering complex random patterns with anisotropic Gaussian transfer functions. The longitudinal and transverse correlation lengths were randomly sampled from morphology-dependent ranges, and a random in-plane rotation was applied to control the orientation of the resulting texture.

We generated waveguide-like patterns to create sparse structure in contrast to the dense blob-like or streak-like patterns. We drew up to three paths per layer by performing Gaussian random walks along the longitudinal direction, with random starting positions along the $x$-axis and step sizes sampled from a discrete Gaussian distribution. Each path was rendered with finite thickness by assigning elevated intensity to neighboring pixels and then smoothed by Gaussian blurring to produce continuous filamentary features. Each layer was assigned a random weight before summation. 

\par We generated 10{,}000 random illumination patterns for the dataset and paired each illumination pattern with 50 independently generated random inputs. The dataset therefore contains 500{,}000 input--output pairs. The inputs and illumination patterns were sent to the experimental setup, and the corresponding one-dimensional output intensity distributions were recorded. Recording the dataset took around 43~hours. Before fine-tuning, the dataset was split into training and validation subsets by illumination pattern (85\% / 15\%), so all 50 input--output pairs for a given pattern belong to the same subset. The validation patterns and their input--output pairs were not used to update the model parameters.

\subsection{Fine-Tuning of Digital Model}
\label{sec:DT fine-tuning}
\par
\begin{figure}[!htbp]
    \centering
    \includegraphics[width=1\linewidth]{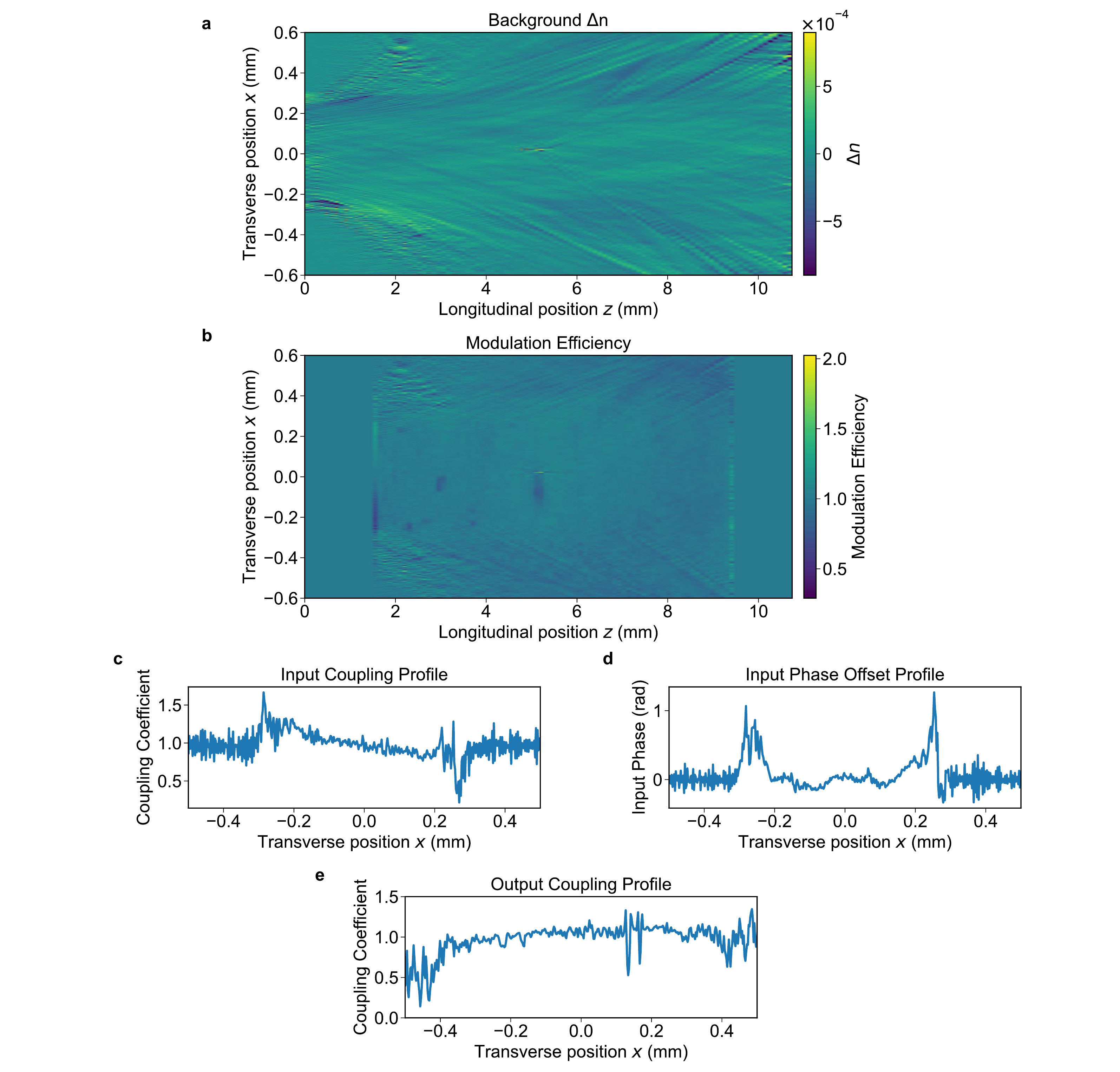}
    \caption{\textbf{Examples of learned effective calibration parameters in the trained model.} \textbf{a}, Effective background-index pattern accounting for static wavefront distortion. \textbf{b}, Effective modulation-efficiency distribution accounting for spatially non-uniform illumination and other deviations from the idealized index-modulation model. \textbf{c}, Effective input-amplitude coupling coefficient accounting for non-uniform amplitude control and coupling at the input facet. \textbf{d}, Effective input-phase coupling coefficient accounting for non-uniform phase control of the free-space beam shaper. \textbf{e}, Effective output coupling coefficient accounting for spatially non-uniform output coupling. These learned quantities collectively capture unmodeled experimental nonuniformities; their individual physical contributions are not independently identified from the intensity-only measurements.}
    \label{fig:DT_param}
\end{figure}

\begin{figure}[!htbp]
    \centering
    \includegraphics[width=1\linewidth]{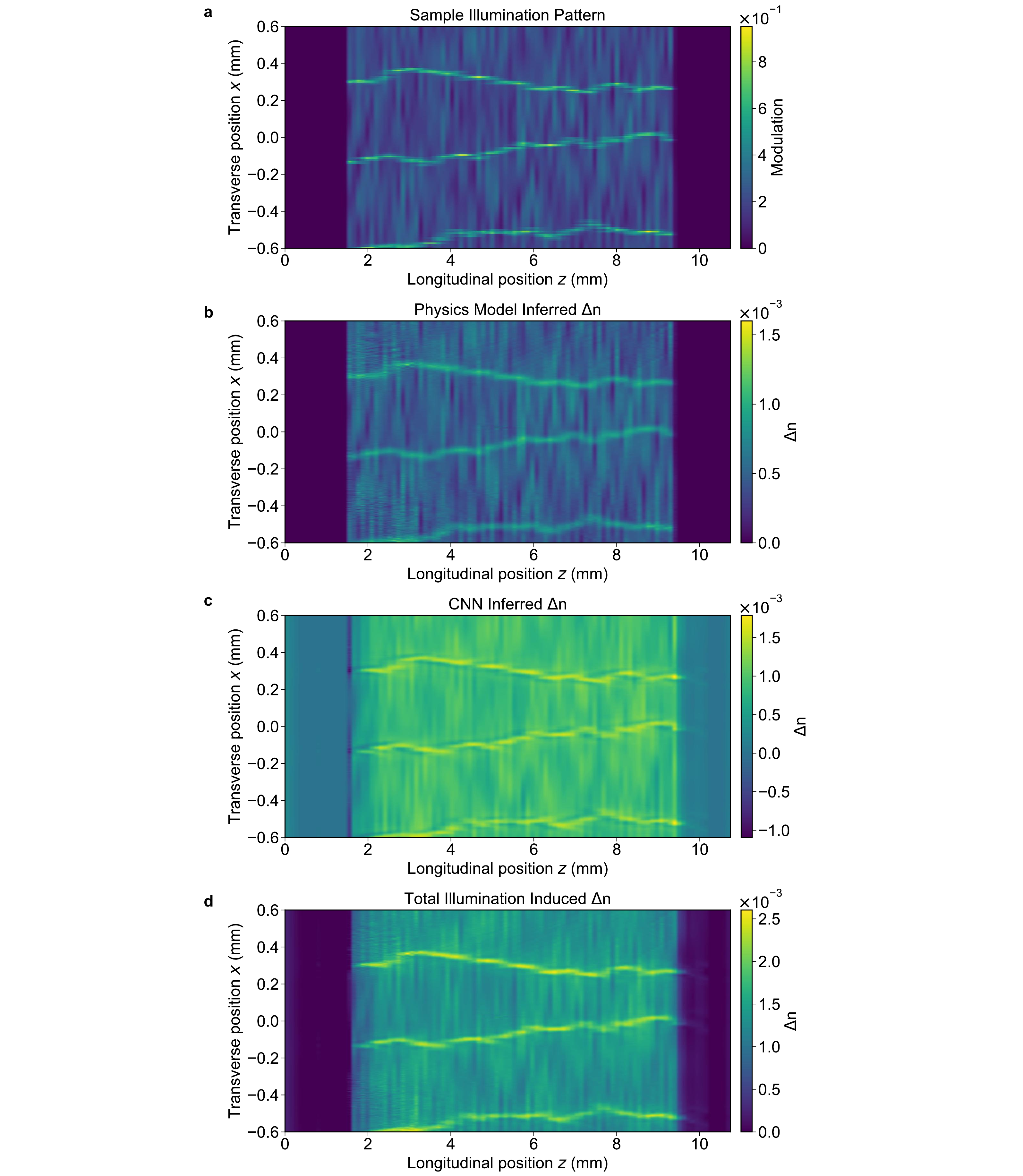}
    \caption{\textbf{Example of the trained model prediction for a random sample pattern.} \textbf{a}, Random DMD image sample sent to the model. \textbf{b}, Inferred index pattern from the physical model. \textbf{c}, Inferred correction to the index pattern from the trained convolutional neural network (CNN). \textbf{d}, Inferred total index-change pattern from the model.}
    \label{fig:DT_sample_prediction}
\end{figure}

\begin{figure}[!htbp]
    \centering
    \includegraphics[width=1\linewidth]{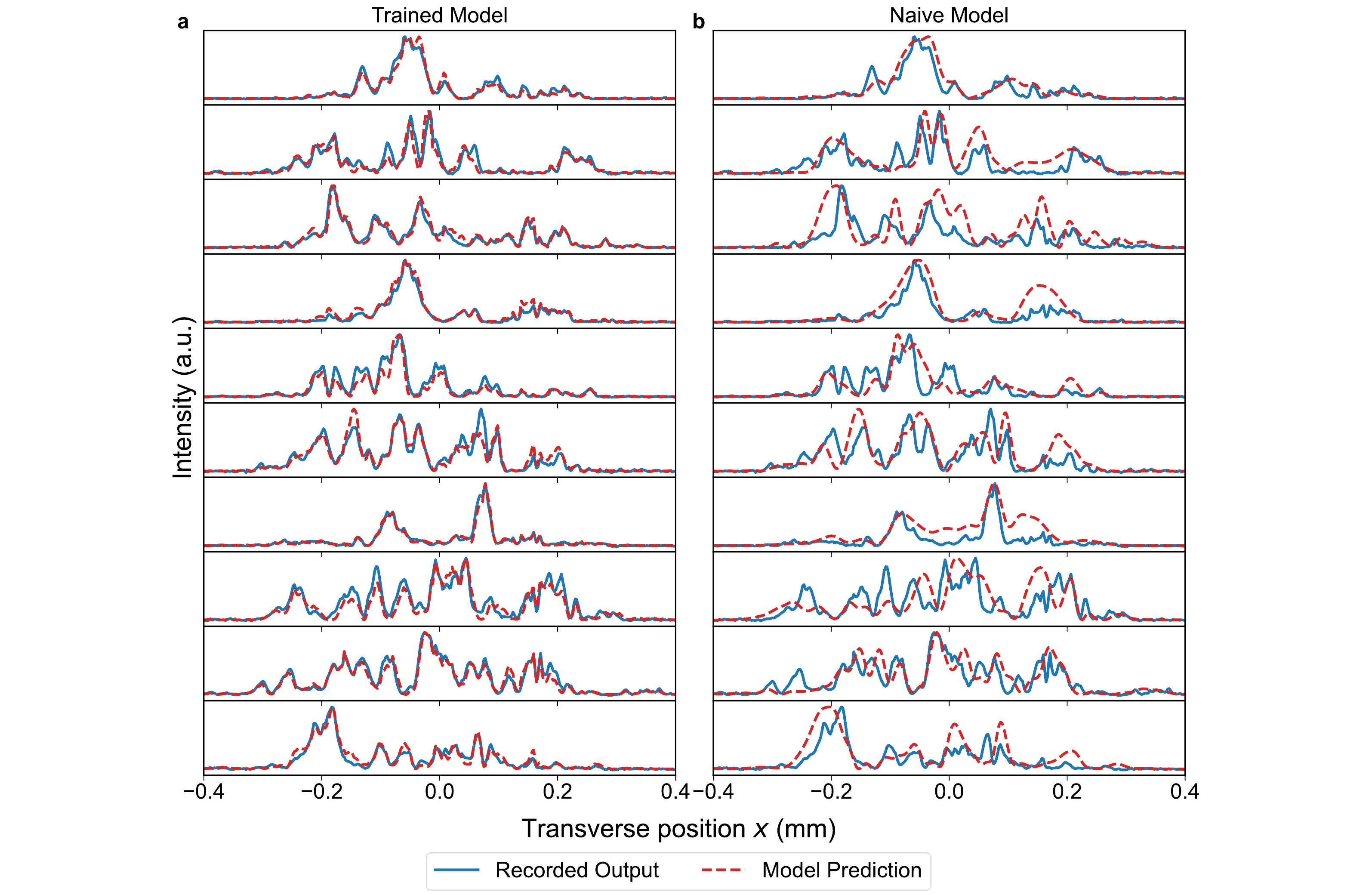}
    \caption{\textbf{Model agreement on representative samples from the held-out validation set.} The illumination patterns and corresponding input--output pairs in this set were not used to update the model parameters. The solid blue curves show the recorded output intensity distributions, and the dashed red curves show the model predictions. \textbf{a}, Predictions from the trained model. \textbf{b}, Predictions from the untrained naive model.}
    \label{fig:DT_agreement}
\end{figure}

 The recorded outputs, along with the random inputs and illumination patterns, were used to fine-tune the digital model. The model has around 467{,}000 parameters, and we fine-tuned it using the following loss function:
\begin{equation}
        \mathcal{L} = 1-
\frac{
\displaystyle \frac{1}{N}\sum_{i=1}^{N}
\left|E_{\mathrm{exp},i}^{*}\!\cdot E_{\mathrm{pred},i}\right|^{2}
}{
\displaystyle
\left(\frac{1}{N}\sum_{i=1}^{N}P_{\mathrm{exp},i}\right)
P_{\mathrm{pred}}
},
\end{equation}
where $E_\text{exp,i}=\sqrt{I_{\text{exp,i}}}$ is the experimental output field derived from recorded output intensity, based on our assumption in Section~\ref{sec:Inverse design process}, $E_{\text{pred,i}}$ is the predicted output field, the sum is over the batch and $N$ is the batch size, $P_\text{exp,i}=\int I_{\text{exp,i}}(x)dx$ is the total power of the recorded output, and $P_\text{pred}=\int I_{\text{pred}}(x)dx$ is the power of the predicted output. We designed the loss function so that the experimental output intensities in each training batch are normalized by the average output power within the batch. This preserves relative power differences between inputs, so that quantities such as the coupling efficiency can be learned. The learning rate was set to $10^{-2}$ with an exponential learning-rate scheduler with $\gamma = 0.99$. Training ran for 25~epochs and took around 80~minutes. Representative learned calibration parameters and the model agreement with the held-out validation data are shown in Fig.~\ref{fig:DT_param}, Fig.~\ref{fig:DT_sample_prediction}, and Fig.~\ref{fig:DT_agreement}.


\subsection{In Situ Inverse Design Process}
\label{sec:PAT}
The model was used in the inverse design to provide the gradient estimate at each step. The training process is largely similar to that of our previous work~\cite{Onodera2025}, which uses physics-aware training~\cite{wrightDeepPhysicalNeural2022}. The inputs are launched sequentially, and the corresponding output intensities are recorded and passed to the training loop. Since in all of our tasks, the input \unitforN{}s share the same power, we take the average power across the output \unitforN{}s and divide each output intensity distribution by this power, so that the mean normalized output power across the input \unitforN{}s is unity while \unitforN{}-to-\unitforN{} power differences are preserved. We chose this normalization because different inputs experience different propagation losses, leading to different intensities at different output \unitforN{}s, and because normalizing by the average power gives a smoother gradient for inverse design. This average-power normalization is used for training only. For the results in Fig.~\ref{fig:2} and Fig.~\ref{fig:3}, we normalize by the maximum output power instead, assuming that the brightest output \unitforN{} experiences coupling and propagation losses similar to those of the others, which preserves the differences between output \unitforN{}s. An additional \SI{5.3}{\decibel} of facet and propagation loss applies (Appendix~\ref{sec:insertion}).

\par The normalized output is then sent to the inverse-design training. To optimize the device toward a desired output spatial mode, we employed a loss function that maximizes the target coupled power and minimizes crosstalk power. Because only intensity is measured, we assume a uniform output phase for the normalized recorded output intensity distribution \(y_i(x)\) of the \(i\)-th input. The output field is then estimated as
\begin{equation}
E^{(i)}_{\mathrm{out}}(x) = \sqrt{\max\!\left(y_i(x), \epsilon\right)},
\end{equation}
where \(\epsilon\) is a small positive constant introduced for numerical stability. The normalized power projected from the measured output onto the desired target mode \(E^{(i)}_{\mathrm{tar}}(x)\) is then written as
\begin{equation}
\eta_i =
\frac{
\left|
\int
\left(E^{(i)}_{\mathrm{out}}(x)\right)^*
E^{(i)}_{\mathrm{tar}}(x)
\, dx
\right|^2
}{
\int
\left|E^{(i)}_{\mathrm{tar}}(x)\right|^2
\, dx
}.
\end{equation}
The quantity \(\eta_i\) therefore corresponds to the normalized power projected onto the desired target mode. The loss function is defined as
\begin{equation}
\mathcal{L}_{\mathrm{ov}} = N - \sum_{i=1}^{N} \eta_i,
\end{equation}
where \(N\) is the batch size. Minimizing \(\mathcal{L}_{\mathrm{ov}}\) is thus equivalent to maximizing the modal overlap between the predicted output and the desired target mode. Compared with a conventional mean-squared-error loss function, this function is better matched to the target optical metrics, as it optimizes relative insertion losses. In our tests, the overlap loss outperformed mean-squared-error (MSE), $L_1$, and cosine-similarity losses. To suppress residual coupling into undesired output modes, we add a crosstalk penalty based on modal overlap. The undesired power is evaluated as
\begin{equation}
\xi_{i,j}
=
\frac{
\left|
\int
\left(E^{(i)}_{\mathrm{out}}(x)\right)^*
E^{(j)}_{\mathrm{tar}}(x)
\, dx
\right|^2
}{
\int
\left|E^{(j)}_{\mathrm{tar}}(x)\right|^2
\, dx
},
\end{equation}
where $j\neq i$ runs over the target modes of other inputs modes. The quantity \(\xi_{i,j}\) represents the normalized power projected onto the target of input \(j\). Rather than summing over all undesired \unitforN{}s, we define the crosstalk penalty using the largest parasitic overlap across the batch:
\begin{equation}
\mathcal{L}_{\mathrm{XT}}
=
\max_{i}\,
\max_{j \neq i}
\xi_{i,j}.
\end{equation}
Minimizing \(\mathcal{L}_{\mathrm{XT}}\) directly suppresses the strongest leakage into the undesired \unitforN{}s, corresponding to a worst-case crosstalk. The total loss is \(\mathcal{L}_{\mathrm{ov}} + 0.6\,\mathcal{L}_{\mathrm{XT}}\).

\par For the power-domain matrix-vector multiplication task, we used a matrix-level loss function based on the normalized mean-squared error (NMSE) between the measured and target modal power-coupling matrices. For each injected input mode \(i\), the normalized output intensity distribution \(y_i(x)\) was first converted into an output field-amplitude estimate \(E^{(i)}_{\mathrm{out}}(x)\) as above. The same procedure is applied to the target intensity distribution to obtain \(E^{(i)}_{\mathrm{tar}}(x)\). The measured and target modal power-coupling matrices are then constructed by projecting each output field-amplitude proxy onto the set of output basis modes \(\phi_j\). The corresponding matrix elements are
\begin{equation}
\left(M_{\mathrm{meas}}\right)_{ji}
=
\frac{
\left|
\int
\left(E^{(i)}_{\mathrm{out}}(x)\right)^*
\phi_{j}(x)
\, dx
\right|^2
}{
\int
\left|\phi_{j}(x)\right|^2
\, dx
},
\end{equation}
and
\begin{equation}
\left(M_{\mathrm{tar}}\right)_{ji}
=
\frac{
\left|
\int
\left(E^{(i)}_{\mathrm{tar}}(x)\right)^*
\phi_{j}(x)
\, dx
\right|^2
}{
\int
\left|\phi_{j}(x)\right|^2
\, dx
}.
\end{equation}
Each matrix element represents the normalized power coupled from input mode \(i\) into output mode \(j\). The loss is defined as the normalized squared Frobenius distance between the two matrices,
\begin{equation}
\mathcal L_{\mathrm{mat}}
=
\frac{
\sum_{i,j}
\left|
(M_{\mathrm{meas}})_{ji}-(M_{\mathrm{tar}})_{ji}
\right|^2
}{
\sum_{i,j}
\left|(M_{\mathrm{tar}})_{ji}\right|^2+\epsilon
}.
\end{equation}
Minimizing \(\mathcal{L}_{\mathrm{mat}}\) drives the experimentally realized modal power-coupling matrix toward the desired target matrix in a relative least-squares sense. 

\subsection{Multi-Wavelength Integration}
\label{sec:multilam}
For the results in Fig.~\ref{fig:3} and Fig.~\ref{fig:4}, we made adjustments to our setup and model to accommodate different wavelengths. First, we adjusted the phase pattern of our SLM-based beam shaper to account for the operating wavelength. Our free-space beam shaper operates on the principle of spatially varying phase gratings on a 2D-phase SLM~\cite{benderDepthtargetedEnergyDelivery2022, frumkerPhaseAmplitudePulse2007}. We adjusted the phase pattern by changing the grating period according to
\begin{equation}
    d = \frac{\lambda}{\lambda_0}d_0,
\end{equation}
where $d$ is the corresponding grating period at wavelength $\lambda$, and $d_0=16$ SLM pixels is the default grating period at $\lambda_0=\SI{1550}{\nano\metre}$. We used the actual wavelength in the back-propagation from the facet to the SLM and rescaled the phase to correct the look-up table calibrated at \SI{1550}{\nano\metre}:
\begin{equation}
    \phi = \frac{\lambda}{\lambda_0}\phi_0,
\end{equation}
where $\phi$ is the 2D phase SLM pattern for wavelength $\lambda$, and $\phi_0$ is the intended 2D phase SLM pattern for the default wavelength $\lambda_0=\SI{1550}{\nano\metre}$. In principle, the most accurate way would be to recalibrate the look-up table for the SLM at each wavelength; however, because the wavelength scan in Fig.~\ref{fig:3} included many wavelengths (\SIrange{1500}{1600}{\nano\metre} in \SI{1}{\nano\metre} steps), we used this approximation instead, which gave satisfactory beam shaping.
\par In the model, we included the material dispersion of LN. At different wavelengths, the dispersion of the LN core changes the effective index of the waveguide stack and the mode confinement. The confinement, in turn, affects the electro-optic modulation efficiency in LN. Both effects influence the programmable index change. The wavenumber and other wavelength-dependent parameters were updated accordingly. We computed all these properties before training, and at each training step, the parameters for the current wavelength were used in the backward pass. In this way, we take into account the different propagation behavior at each wavelength.
\par The C-band laser is continuously tunable, enabling the wavelength scans from \SI{1500}{\nano\metre} to \SI{1600}{\nano\metre} shown in Fig.~\ref{fig:3}. In contrast, the \SI{1310}{\nano\metre} source used for Fig.~\ref{fig:4} operates at a fixed wavelength. We therefore evaluated the wavelength-demultiplexing device at the two discrete design wavelengths of \SI{1310}{\nano\metre} and \SI{1550}{\nano\metre}.

\subsection{Resilience of in situ training to larger simulation--reality gaps}
\label{sec:PAT_advantage}

In this work, the main technique we used to perform in situ inverse design of our programmable device is physics-aware training~\cite{wrightDeepPhysicalNeural2022}. Unlike conventional inverse design or training, where optimization of the refractive index pattern is performed purely through digital simulation, physics-aware training replaces the digital forward pass with a physical, experimental forward pass. This allows the training routine to integrate experimental measurements into gradient updates to counteract the simulation--reality gap. This is central to our approach, enabling us to narrow the simulation--reality gap on this programmable device more effectively than in conventional inverse-designed devices.

\begin{figure}[!htbp]
    \centering
    \includegraphics[width=1\linewidth]{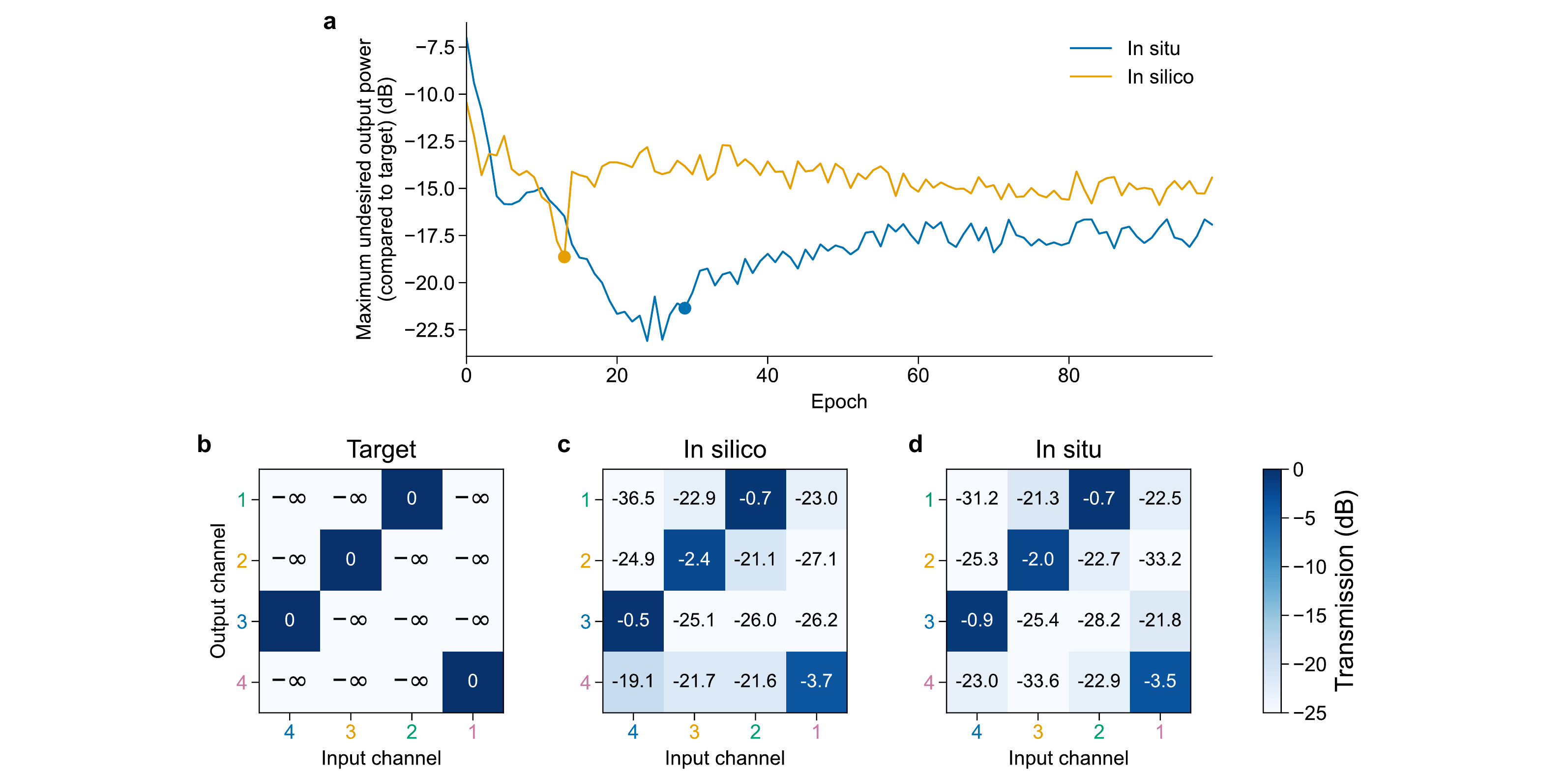}
    \caption{\textbf{Measured advantage from in situ training.} \textbf{a}, Maximum undesired output power from experiment during training compared to the target power, showing both results using in silico training and physics-aware training. \textbf{b}, Target transmission matrix for $4\times4$ permutation task. \textbf{c}, Measured best transmission matrix for in silico training. \textbf{d}, Measured best transmission matrix for in situ training. }
    \label{fig:PAT_advantage}
\end{figure}

To illustrate the benefits of physics-aware training, we reproduced a digital model with our best effort, and the model showed a slightly lower accuracy. We repeated the inverse design of the $4\times4$ permutation task demonstrated in Section~\ref{sec:Single_wavelength}, comparing physics-aware training with in silico training (with the optimization carried out purely using a digital model, but passively monitoring the device performance at each training iteration). The results are shown in Fig.~\ref{fig:PAT_advantage}. We achieved a crosstalk of approximately \SI{-20}{\decibel} using physics-aware training, whereas in silico training yielded a worse crosstalk of around \SI{-15}{\decibel}. This test illustrates the advantage of in situ training in the case of a worse simulation--reality gap: it mitigates the gap using the physical forward pass and a real-time update of the pattern, leading to improved performance.

\section{Future Device Improvements}
\label{sec:improvements}
\subsection{Future Improvements for Our Current Device}
This appendix discusses possible improvements to the present device. Here we use the notation of Appendix~\ref{sec:design}. As noted in Appendix~\ref{sec:design}, $\mathcal{R}$ is field dependent.
\label{sec: Future Device}
\subsubsection{Photoconductor}
Some of the key performance parameters of our device are constrained by the photoconductor material.
When illuminating the photoconductor with photons of energy greater than its bandgap, the absorbed photons generate free carriers. The generation rate of carriers per unit volume takes the form~\cite{seeger2013semiconductor}
\begin{align}
\label{eq:generation-rate}
G=\frac{\alpha \eta I_\text{opt}}{\hbar \omega},
\end{align}
where $\alpha$ is the absorption coefficient of the photoconductor, $\eta$ is the quantum efficiency (the number of carriers generated per absorbed photon), $\hbar\omega$ is the energy per photon, and $I_\text{opt}$ is the intensity of the programming illumination. Here, we assume that the intensity of the programming illumination stays roughly constant regardless of the distance from the surface. This condition requires that the thickness of the photoconductive layer is smaller than the characteristic absorption length $\alpha^{-1}$.

The steady-state photo-induced excess carrier density $\Delta N_\mathrm{c}$ is given by the balance between the generation and decay rates as $\Delta N_\mathrm{c}=G\tau$, where we assume equal excess electron and hole densities and the same carrier lifetime $\tau$ for both carrier types. The corresponding change in conductivity takes the form~\cite{seeger2013semiconductor}
\begin{align}
    \Delta \sigma=\Delta N_\mathrm{c} q(\mu_n+\mu_p)=G\tau q(\mu_n+\mu_p),
\end{align}
with electron charge $q=\SI{1.602e-19}{\coulomb}$, and the electron and hole mobilities $\mu_n$ and $\mu_p$, respectively.

For our devices, the relevant figure of merit is the photoconductive switching ratio
\begin{align}
    \mathcal{R}=\frac{\Delta \sigma}{\sigma_0},
\end{align}
where $\sigma_0$ is the intrinsic conductivity of the photoconductor with no programming illumination. There are several design principles to maximize the figure of merit $\mathcal{R}$. First, the wavelength of the programming illumination so that $\alpha$ satisfies $d\alpha\sim 1$ for the photoconductive layer thickness $d$, because this ensures that the incoming photons are efficiently utilized without significant transmission through the film, while the film is not so thick that only a thin surface layer is activated.
Second, we should use a photoconductor with a long carrier lifetime $\tau$. This is the opposite of the requirement for fast photoconductive switches, e.g., in terahertz generation. Finally, the dark carrier density of the photoconductor should be low to minimize $\sigma_0$.

As a case study, we consider intrinsic crystalline silicon (c-Si). At a wavelength of \SI{780}{\nano\metre}, with a photon energy of $\hbar \omega=\SI{2.6e-19}{\joule}$, the absorption coefficient of c-Si is approximately $\alpha=\SI{1e3}{\per\centi\metre}$, corresponding to a penetration depth of \SI{10}{\micro\metre}~\cite{sze2021semiconductor}. For the photoconductor thickness $d=\SI{6}{\micro\metre}$ of our device, the approximation in Eq.~\eqref{eq:generation-rate} is reasonable. With $I_\text{opt}=\SI{1}{\micro\watt\per\square\centi\metre}$ of programming illumination, the carrier generation rate per volume becomes $G=\SI{4e15}{\per\cubic\centi\metre\per\second}$ at $\eta\sim 1$.
For a typical carrier lifetime of $\tau=\SI{1}{\milli\second}$~\cite{richter2012carrierlifetime}, this corresponds to the steady-state carrier density of $G\tau=\SI{4e12}{\per\cubic\centi\metre}$. With typical carrier mobilities of $\mu_n=\SI{1430}{\centi\metre\squared\per\volt\per\second}$~\cite{li1977electronmobility} and $\mu_p=\SI{450}{\centi\metre\squared\per\volt\per\second}$~\cite{jacoboni1977holemobility} at \SI{300}{\kelvin}, we obtain a photoconductivity of $\Delta \sigma=\SI{1e-3}{\per\ohm\per\centi\metre}$. Highly insulating c-Si can exhibit conductivities as low as $\sigma_0=\SI{3.3e-6}{\per\ohm\per\centi\metre}$~\cite{green1990intrinsicSi}, corresponding to a photoconductive switching ratio of $\mathcal{R}=300$ with $I_\text{opt}=\SI{1}{\micro\watt\per\square\centi\metre}$ of illumination intensity.

These numbers indicate that a substantially higher photoconductive switching ratio is achievable with an appropriate photoconductor material, far in excess of what is needed for functional programmable photonic devices. With an optimized or alternative photoconductor, the following improvements become possible.

\subsubsection{Potential improvements with a better photoconductor}
\par \textbf{Speed:} There are two distinct effects describing the speed of our device: (i) the recording speed for different inputs, and (ii) the switching speed for different illumination patterns. The former is constrained by our recording and beam shaping setup, and can be improved by using high-speed modulation on the input and high-speed recording on the output. Although we can only record when the voltage is at maximum, if the modulation and recording are much faster than the AC frequency, we can still record multiple input--output pairs around the maximum, and continue in the next cycle. In this case we can decouple input and output recording speed from the AC voltage, and the recording speed is not limited by the device.

We now discuss the switching speed of the pattern. The switching speed of our device is constrained by the RC time constant of the photoconductor--waveguide stack. Even if the photoconductor switches instantaneously, the field in the LN takes ${\sim}RC$ to settle, resulting in limited pattern switching speed. In principle, if we can make the photoconductor layer more conductive, the pattern switching speed could be made much faster than that shown in this work.

We modeled the field transition process when switching between the bright and dark states as follows. The electrical response of the chip can be characterized as
\begin{equation}
C_{\Sigma}\,\frac{\mathrm{d}V_{\mathrm{wg}}}{\mathrm{d}t} = G_{\mathrm{pc}}\left(V - V_{\mathrm{wg}}\right) + C_{\mathrm{pc}}\,\frac{\mathrm{d}V}{\mathrm{d}t}
\end{equation}
and the index modulation follows instantaneously through the Pockels effect, $\Delta n(t) = K V_{\mathrm{wg}}(t)$. Illumination changes the conductance of the photoconductor between its dark value $G_{\mathrm{d}}$ and bright value $G_{\mathrm{b}} = (1+\mathcal{R})G_{\mathrm{d}}$, where $\mathcal{R}$ is the switching ratio. For a sinusoidal AC voltage $V(t) = V_{0}\sin\omega t$, the steady state is $V_{\mathrm{wg}}(t) = \operatorname{Im}[H(\omega,G_{\mathrm{pc}})V_{0}e^{j\omega t}]$ with $H(\omega,G_{\mathrm{pc}}) = (G_{\mathrm{pc}} + j\omega C_{\mathrm{pc}})/(G_{\mathrm{pc}} + j\omega C_{\Sigma})$. Because the equation is first-order and linear, the response to a change of illumination at time $t_{s}$ is
\begin{equation}
V_{\mathrm{wg}}(t) = V_{\mathrm{wg}}^{\mathrm{ss}}(t) + \left[V_{\mathrm{wg}}(t_{s}) - V_{\mathrm{wg}}^{\mathrm{ss}}(t_{s})\right]e^{-(t-t_{s})/\tau}, \qquad \tau = C_{\Sigma}/G_{\mathrm{new}},
\end{equation}
where $V_{\mathrm{wg}}^{\mathrm{ss}}$ is the steady state of the new illumination state and $G_{\mathrm{new}}$ is the photoconductor conductance in the new state. The pattern switching speed is set by the conductance the pixel switches \emph{to}, with $\tau_{\mathrm{on}} = C_{\Sigma}/G_{\mathrm{b}} \approx \SI{22}{\milli\second}$ and $\tau_{\mathrm{off}} = C_{\Sigma}/G_{\mathrm{d}} = (1+\mathcal{R})\tau_{\mathrm{on}} \approx \SI{300}{\milli\second}$ for the fitted device parameters in Appendix~\ref{sec:holography}. When a pattern is updated, the bright-to-dark and dark-to-bright transitions occur simultaneously. The normalized contrast between an on pixel and an off pixel then evolves as $c(t)/c_{\infty} = 1 - e^{-t/\tau_{\mathrm{on}}} - e^{-t/\tau_{\mathrm{off}}}$, so the update time is dominated by $\tau_{\mathrm{off}}$. The time to reach 99\% of index change is calculated to be \SI{1.4}{\second}, consistent with the experimentally observed \SI{1.5}{\second} switching time. If we make the dark state conductivity of the photoconductor 100 times larger, then the response time could be on the order of \si{\milli\second}. However, this would result in higher electrical power consumption, as discussed below.

\par \textbf{Resolution:} At present, the spatial resolution of the programmable refractive-index profile is mostly limited by the pixel size of the DMD and the resolution of the optical imaging system used to project the illumination pattern onto the chip. In the current setup, these factors result in an effective minimum pixel size of approximately \SI{9}{\micro\metre}. This limitation is not fundamental to the device concept itself, and in principle could be improved substantially using a higher-resolution DMD (e.g., 4K) and higher-NA projection optics.

Once these factors are improved, however, the spatial resolution of the programmable index profile will ultimately be constrained by the electric-field fringing across the multilayer stack between the top photoconductive layer and the lithium niobate core of the waveguide. When the field for a single bright pixel extends across multiple layers, the field will spread farther away, leading to a further smoothing of the electric field profile in the LN core. To characterize this effect experimentally, we projected a very narrow inverse-waveguide feature, i.e., a strip with a lower index than its surroundings, only one or two DMD pixels wide. We launched an input beam at an angle such that the strip partially reflected it. Because the sharpness of the index profile at the strip edges changes how the power divides between the reflected and transmitted beams, we could extract the edge sharpness from this split, assuming that the uniform index outside the strip is known and fixed. We therefore extracted an effective fringing width of approximately \SI{3}{\micro\metre}.

This fringing scale is set primarily by the geometry of the device stack and the separation between the top electrode, photoconductive layer, and the lithium niobate core, and therefore cannot be overcome by improving the external imaging optics. These results suggest that, although the present \SI{9}{\micro\metre} resolution can in principle be improved substantially, the ultimate resolution of the programmable index profile will likely be bounded by the fringing length on the order of a few \si{\micro\metre}. 

\par \textbf{Maximum Index Change Capability:} The maximum index change is set by the maximum field contrast inside the LN core between the dark state and bright state. From Eq.~\ref{eq:delta_n_max}, at fixed voltage and capacitances, $\Delta n_{\mathrm{prog,max}}$ depends only on $\mathcal{R}$. Another limiting factor is the breakdown field of all layers, and the higher the voltage the device can reach, the more index change it can achieve. For an ideal photoconductor ($\mathcal{R}\gg1$, negligible capacitance $C_{\mathrm{pc}}\sim0$), the maximum index change is set by the thin-film LN breakdown field (\SI{65}{\kilo\volt/\milli\metre}~\cite{luennemannElectroopticPropertiesLithium2003}), giving $\Delta n\approx10^{-2}$. However, the AC driving scheme we used might be limited by the coercive field of LN, which can vary between samples. Using the coercive field reported for thin-film LN (\SI{34}{\kilo\volt/\milli\metre}~\cite{nagySubmicrometerPeriodicPoling2020}), $\Delta n$ is limited to ${\sim}5\times10^{-3}$. The index modulation will be limited by the capacitance of the photoconductor as well, which limits $\Delta n$ to ${\sim}2\times10^{-3}$ for the present chip. Reducing $C_{\mathrm{pc}}$ (e.g., via a thicker or lower-permittivity photoconductor) would raise this limit.

\par \textbf{Power Consumption:} The power consumption of our device comes from two parts: the illumination power needed to maintain a conductive state in the photoconductor and induce the index change, and the high voltage needed to maintain the electric field for the EO effect in LN. We currently measure the illumination power on top of the chip to be around \SI{3}{\milli\watt}, while the electrical power drawn from the high-voltage supply is around \SI{2}{\milli\watt}. The large amount of power used for illumination is mostly for activating the SRN layer, which is a comparatively inefficient photoconductor. A better photoconductor would provide more than enough switching (see the c-Si case study above). Compared with the maximum intensity of $I_\text{opt}=\SI{50}{\milli\watt\per\square\centi\metre}$ used in the present experiment, the c-Si case study suggests that an appropriate photoconductive material could reduce the illumination intensity, and hence the illumination power, by about four orders of magnitude, from about \SI{3}{\milli\watt} to the order of \SI{1}{\micro\watt}, although developing the required fabrication techniques remains challenging.
\par We next consider the electrical power drawn from the high-voltage source. In the circuit model discussed in Appendix~\ref{sec:design}, our device can be represented by a photoconductive layer in series with the electro-optic waveguide stack, which is modeled as a capacitor. With \(V_{\mathrm{app}}\) the peak amplitude of the sinusoidal applied voltage, the time-averaged power drawn from the source is
\begin{equation}
P_{\mathrm{HV}}
=
\frac{1}{2}V_{\mathrm{app}}^{2}
\frac{G_{\mathrm{pc}}\omega^{2}C_{\mathrm{wg}}^{2}}
{G_{\mathrm{pc}}^{2}+\omega^{2}(C_{\mathrm{pc}}+C_{\mathrm{wg}})^{2}}.
\label{eq:power}
\end{equation}
where $G_{\mathrm{pc}}$ is the conductance of the photoconductor under a given illumination power, such that $G_{\mathrm{d}}<G_{\mathrm{pc}}<G_{\mathrm{b}}$. This expression shows that the high-voltage power is determined by the photoconductor conductance and the capacitance of the stack. In particular, \(P_{\mathrm{HV}}\) decreases when increasing \(C_{\mathrm{pc}}\), whereas its dependence on \(G_{\mathrm{pc}}\) is non-monotonic, reaching a maximum at
\begin{equation}
G_{\mathrm{pc}}=\omega(C_{\mathrm{pc}}+C_{\mathrm{wg}}).
\end{equation}
At the optimal frequency (Eq.~\ref{eq:opt_freq}), the power becomes
\begin{equation}
P_{\mathrm{HV}}^{\mathrm{opt}}(G_{\mathrm{pc}})
= \frac{1}{2} V_{\mathrm{app}}^{2}
  \left(\frac{C_{\mathrm{wg}}}{C_{\Sigma}}\right)^{2}
  \frac{G_{\mathrm{pc}}\,G_{\mathrm{b}} G_{\mathrm{d}}}
       {G_{\mathrm{pc}}^{2} + G_{\mathrm{b}} G_{\mathrm{d}}}
= \frac{1}{2} V_{\mathrm{app}}^{2}
  \left(\frac{C_{\mathrm{wg}}}{C_{\Sigma}}\right)^{2}
  G_{\mathrm{d}}\,
  \frac{(1+\mathcal{R})\,g}{g^{2} + 1 + \mathcal{R}}.
  \label{eq:power_opt}
\end{equation}
where $g \equiv G_{\mathrm{pc}}/G_{\mathrm{d}}$ and $C_\Sigma=(C_{\text{wg}}+C_{\text{pc}})$. This power consumption is the same for the brightest and darkest pixels, but will reach a maximum at $g=\sqrt{1+\mathcal{R}}$, with power of
\begin{equation}
    P_{\mathrm{HV,max}}^{\mathrm{opt}}
= \frac{1}{4} V_{\mathrm{app}}^{2}
  \left(\frac{C_{\mathrm{wg}}}{C_{\Sigma}}\right)^{2}
  G_{\mathrm{d}}\,\sqrt{1+\mathcal{R}}=\frac{1}{4} V_{\mathrm{app}}^{2}
  \left(\frac{C_{\mathrm{wg}}}{C_{\Sigma}}\right)^{2}\sqrt{G_{\mathrm{d}}G_{\mathrm{b}}}
  \label{eq:power_opt_worst}
\end{equation}
For the fitted $\mathcal{R}$, this is about twice the power consumption of a dark pixel. Assuming a worst-case condition for all pixels, the power consumption is calculated to be around \SI{1}{\milli\watt}. This value applies at the optimal frequency of about \SI{2}{\hertz}. For our experimental condition of \SI{4}{\hertz}, the worst-case power consumption is calculated to be around \SI{1.8}{\milli\watt}. We measured the electrical consumption under uniform illumination at 50\% maximum intensity by measuring the total current received by the voltage source. For an applied peak voltage of $V_{\mathrm{app}}=\SI{600}{\volt}$, we measured a peak current of $I_{\mathrm{pk}}=\SI{25}{\micro\ampere}$, and voltage--current phase difference $\phi\approx0.4\pi$. Therefore we measured a dissipated power of $P_{\mathrm{HV,exp}}=\tfrac{1}{2}V_{\mathrm{app}}I_{\mathrm{pk}}\cos\phi\approx\SI{2.3}{\milli\watt}$, reported here as approximately \SI{2}{\milli\watt}. Some portion of it may also be attributed to other resistances in the circuits, such as the power source. Equation~\ref{eq:power_opt_worst} shows that the power consumption will increase as we increase the dark conductivity of the photoconductor. A 100-fold increase in $G_{\mathrm{d}}$ (as above) would raise this to about \SI{200}{\milli\watt} at fixed $\mathcal{R}$. This creates a trade-off between power consumption and pattern switching speed.

\subsubsection{Potential quasi-DC operation}
\label{sec:flickering}
The AC drive scheme has two drawbacks. In a full AC cycle, we can only record at the specific time where the voltage in the lithium niobate core reaches maximum, which could introduce many compromises. For example, most of each AC period is unused, spent waiting for the voltage to rise or for the waveguide capacitor to charge. In addition, the stack capacitance is charged and discharged every cycle, dissipating energy. We therefore propose synchronizing the illumination pattern with the applied AC voltage to avoid unnecessary discharge of the capacitor.

In this technique, the illumination pattern cycles through three frames per voltage period: the positive pattern near the positive voltage peak, the negative pattern near the negative peak, and a dark frame in between. Consider a pixel for which we want a positive index change. The illumination is projected on the chip only when the voltage is over a threshold voltage $V_{\text{th}}$, around the peak of the positive AC voltage. In this case, the waveguide capacitor is charged only near the voltage peak and keeps the voltage $V_{\text{th}}$ at which the illumination is switched off, so tuning $V_{\text{th}}$ up to $V_{0}$ sets the index change continuously, with the maximum as $V_{\text{th}}$ approaches $V_{0}$. Between peaks, the pixel slowly discharges through $G_{\mathrm{d}}$, so its voltage settles at a quasi-steady (``quasi-DC'') value near the threshold voltage. Likewise, pixels illuminated only near the negative voltage peak acquire a negative index change, so the index change is no longer bounded below by zero. While a pixel is illuminated, its photoconductor becomes more conductive and the waveguide voltage charges at the rate $G_{\mathrm{b}}/C_{\Sigma}$; when the illumination is removed, the pixel discharges slowly through $G_{\mathrm{d}}$, at the rate $G_{\mathrm{d}}/C_{\Sigma}$. Each pixel therefore maintains a dynamic balance at the charged voltage. With $\tau_{\mathrm{on}}$ and $\tau_{\mathrm{off}}$ as defined above, the charging and discharging of the pixel in the operating window $\tau_{\mathrm{on}} \ll 1/f \ll \tau_{\mathrm{off}}$ could be described as
\begin{equation}
V_{\mathrm{wg}}(t) = \pm V_{\mathrm{th}} + \kappa\left[V(t) \mp V_{\mathrm{th}}\right],
\end{equation}
where $\kappa = C_{\mathrm{pc}}/C_{\Sigma}$, $V_{\mathrm{th}}$ is the threshold voltage, $V(t)$ is the AC voltage, and the upper (lower) sign applies to positive (negative) pixels. The oscillatory term $\kappa V(t)$ is common to both kinds of pixels, so the programmed contrast $\Delta V_{\mathrm{wg}} = 2(C_{\mathrm{wg}}/C_{\Sigma})V_{\mathrm{th}}$, and hence $\Delta n = K\,\Delta V_{\mathrm{wg}}$, is time-independent. Reprogramming uses the same mechanism: to update the pattern, the applied voltage is set to zero while all pixels are illuminated, so that all waveguide capacitors discharge through the bright conductance, after which the new positive and negative patterns are projected around the subsequent voltage peaks to write the new state.

\subsubsection{Further improvements from the quasi-DC technique}
\par \textbf{Speed:} 
In the quasi-DC scheme, the dynamics are different from those of our conventional AC approach. Each pixel discharges much more slowly when the illumination is off, effectively staying at the threshold voltage. In this case, there will be no asymmetric switching of the on and off state. All switching is performed through the \emph{bright} conductance. When switching patterns, the voltage is set to zero for a short period of time, and all pixels are illuminated so that they discharge quickly. The calculated time to 99\% discharge for the fitted parameters is ${\sim}\SI{0.1}{\second}$, about $10\times$ faster than passive relaxation, because both charging and discharging occur through $G_{\mathrm{b}}$. Increasing $G_{\mathrm{b}}$ therefore shortens both switching times ($\propto 1/G_{\mathrm{b}}$). A $100\times$ increase in $G_{\mathrm{b}}$ would bring the full update time to the millisecond scale, meaning a pattern switching speed as fast as \SI{1}{\kilo\hertz}. However, the \SI{1}{\milli\second} carrier lifetime mentioned above may limit the pattern switching speed beyond this. Nevertheless, the roles of the two conductances are decoupled: $G_{\mathrm{d}}$ determines only the holding fidelity and static dissipation, while $G_{\mathrm{b}}$ alone determines how fast the device can be reprogrammed. 

\par \textbf{Maximum Index Change Capability:} Under static illumination, the contrast between bright and dark pixels is set by
\begin{equation}
\Delta V_{\mathrm{wg}}^{\mathrm{const}} \le \frac{\mathcal{R}}{2+\mathcal{R}}\left(1-\kappa\right)V_{0} < \left(1-\kappa\right)V_{0},
\end{equation}
where the final term is for the case of an ideal photoconductor ($\mathcal{R} \to \infty$). Moreover, illumination only drives pixels toward $V_{0}$, so the pattern is confined to the range $[0, V_{0}]$. However, the quasi-DC scheme goes beyond this bound, because the positive and negative pixels are illuminated at different peaks of the AC voltage. Positive pixels will be activated around $+V_{0}$ and negative pixels will be activated around $-V_{0}$, so the full voltage range will be $[-V_{0}, +V_{0}]$ and the contrast becomes $\Delta V_{\mathrm{wg}} = 2\left(1-\kappa\right)V_{0},$ and the corresponding index contrast is $\Delta n = 2K(1-\kappa)V_{0}$. This is obtained without any change to the device itself, only by changing the illumination sequence. For the fitted device parameters this amounts to $\Delta n \approx 4.2\times10^{-3}$ at $V_{0} = \SI{600}{\volt}$, compared with $1.7\times10^{-3}$ under constant illumination at the same bias. Ideally, the quasi-DC scheme would boost the maximum index change to around $2\times10^{-2}$, but the LN coercive field would again limit the maximum contrast to around $1.5\times10^{-2}$, which is still three times as large as in the conventional AC scheme.

\par \textbf{Power Consumption:} 
An illuminated pixel consumes little power because the voltage across the conductive photoconductor is small; the power consumption is dominated by dark-state leakage. The minimum-index and maximum-index pixels, held at $\pm V_{0}$, dissipate the most, so the maintenance power density is at most
\begin{equation}
P_{\mathrm{quasi\text{-}DC}} = \frac{3}{2}\,G_{\mathrm{d}}\left(\frac{C_{\mathrm{wg}}}{C_{\Sigma}}\right)^{2} V_{0}^{2},
\end{equation}
For a given stack and voltage, the maintenance power therefore scales only with $G_{\mathrm{d}}$. For the current device this power is still comparable to the electrical power consumption of constant illumination (${\sim}\SI{1}{\milli\watt\per\square\centi\meter}$ at \SI{600}{\volt}), but it decreases when we decrease $G_{\mathrm{d}}$. Reducing $G_{\mathrm{d}}$ by a factor of $10^{3}$ would lower the holding power to ${\sim}\SI{1}{\micro\watt\per\square\centi\meter}$, whereas in the conventional AC scheme the electrical power, which is partly determined by the bright pixels, would remain at the milliwatt level. Under such improvements, the dominant electrical energy cost is no longer maintaining a pattern but writing a pattern. Each update charges and discharges the capacitor at an energy cost of $\tfrac{1}{2}C_{\mathrm{wg}}V_{0}^{2}$ (${\sim}\SI{0.3}{\milli\joule\per\square\centi\meter}$ at \SI{600}{\volt}), and it is consumed once per reprogramming.

\subsection{Potential Device Under DC Voltage}
\label{sec:DC device}
\begin{figure}[!htbp]
    \centering
    \includegraphics[width=1\linewidth]{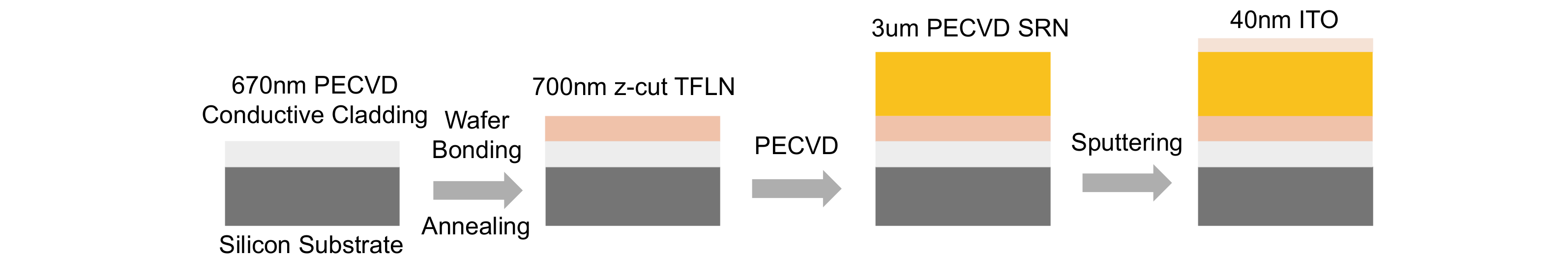}
    \caption{\textbf{Schematic of the fabrication process of the proposed DC-driven device.} The overall process is similar to that of the device in this work, except that the bottom cladding is replaced with a more conductive cladding, and the top cladding is removed to make DC operation possible.}
    \label{fig:DC_device_fab}
\end{figure}

As we discussed in Appendix~\ref{sec: Future Device}, under AC drive the attainable $\Delta n$ is ultimately limited by the coercive field of LN. We can bypass such a limit by setting the device to operate under DC, but the thick and insulating SiO$_2$ screens the field, leaving little field in the LN core. We therefore propose a modified stack for DC operation. The DC-programmed lithium-niobate device considered here operates on the same principle as our 2D programmable device, which uses a photoconductive layer to redistribute the electric field across a z-cut LN film and therefore generate a spatially programmable refractive-index profile through the electro-optic effect. However, the two devices are substantially different in structure. From the bottom to the top, the proposed DC device comprises a p$^{+}$ Si substrate, a conductive boron-doped SiO$_x$N$_y$ bottom layer, a \SI{700}{\nano\metre}-thick MgO-doped z-cut LN film, a \SI{3}{\micro\metre}-thick SRN photoconductor, and a \SI{40}{\nano\metre} ITO top electrode. Fabrication starts from RCA cleaning of the Si wafer, PECVD deposition and polishing of the boron-doped SiO$_x$N$_y$, high-temperature annealing, bonding of the z-cut LN layer, PECVD deposition of the SRN photoconductor, and final ITO deposition. A schematic of the fabrication process for the potential DC device is shown in Fig.~\ref{fig:DC_device_fab}.

The DC device is designed for an optical wavelength of \SI{800}{\nano\metre} because the bottom SiO$_x$N$_y$ layer is too thin to prevent substrate leakage at \SI{1550}{\nano\metre}. Simulations indicate a low substrate loss at \SI{800}{\nano\metre} and a modulation efficiency over 90\% for the chosen \SI{700}{\nano\metre} LN thickness. The expected operating voltage is \SIrange{500}{600}{\volt}, and the modeled index change reaches $\Delta n \sim 10^{-2}$. The DC architecture offers the following advantages over our current design.

\par \textbf{Speed:}
This device operates at DC, so the optical inputs and the camera shots at the output no longer need to be synchronized with the AC driving voltage. The input--output recording speed is then not limited by the device, as for the AC device (Table~\ref{tab:performance_comparison}). However, the switching speed is still limited by the conductivity of the photoconductor, and our simulation also shows a \SI{1}{\second} pattern switching time.

\par\textbf{Maximum Index Change Capability:}
The main advantage of the DC device is its potentially larger programmable refractive-index change. Simulations for the DC device indicate $\Delta n \sim 10^{-2}$ for the designed \SI{700}{\nano\metre}-thick LN layer at \SI{800}{\nano\metre}, with the present photoconductor. We can also raise the field in the LN core above its coercive field, because a DC field parallel to the spontaneous polarization does not reverse the domains, and unlike in our current device, the polarity of the applied field does not need to flip in operation. The DC device can therefore provide substantially larger index modulation than the present AC device. Such a large accessible index change is beneficial for realizing larger-scale devices and smaller chip footprints.

\par\textbf{Field-Fringing Resolution:}
A further advantage of the DC device is the improved spatial resolution of the electric-field distribution due to its simpler vertical structure and the reduced number of material layers. Because the DC stack has fewer layers between the electrode and the LN, lateral field spreading is reduced, shrinking the fringing-induced blurring of the programmed index profile from \SI{3}{\micro\metre} to the sub-\si{\micro\metre} level. This improved control over field fringing would increase the effective resolution of the programmed index profile.

\par\textbf{Power Consumption:} The power consumption for this device is expected to be comparable to that of the present device, since the illumination power and supply current are similar.

\subsection{Potential Device with Electro-Optic Polymer Core}
\label{sec: Polymer Device}
As a further extension of the 2D programmable photonic device, the LN core can be replaced by other electro-optic materials. For example, we could use a poled electro-optic polymer, which has a stronger electro-optic coefficient than lithium niobate. The proposed stack consists of a p-type silicon substrate, a poled electro-optic polymer core, spin-on-glass cladding on both sides, an SRN photoconductor, and a transparent ITO top electrode. The working principle is the same as that of our current AC device, except that the core material has an electro-optic coefficient that could be seven times that of lithium niobate. Ref.~\cite{luHightemperatureresistantSiliconpolymerHybrid2020} reported $n^3r_{33}\approx \SI{1021}{\pico\metre\per\volt}$ for a side-chain EO polymer, and Ref.~\cite{kieningerUltrahighElectroopticActivity2018} reported an $n^3r_{33}\approx \SI{2300}{\pico\metre\per\volt}$ for the EO chromophore JRD1. This would yield $\Delta n$ up to ${\sim}1.5\times10^{-2}$ in the current AC geometry. EO polymers could also tolerate a higher breakdown field than LN (around $\SI{100}{\volt\per\micro\metre}$ in Ref.~\cite{kozanecka2011electro}), so the achievable $\Delta n$ could be even higher. For fabrication, SRN requires a different deposition method from PECVD since the process temperature at $\SI{350}{\celsius}$ is far above the polymer's glass transition temperature ($\SI{172}{\celsius}$ in Ref.~\cite{luHightemperatureresistantSiliconpolymerHybrid2020}). Therefore, we would need to use room-temperature sputtering for SRN deposition. 

Combining the polymer core with the DC stack of Appendix~\ref{sec:DC device}, in which the top cladding is removed, would improve the device further. An index change of $5\times10^{-2}$ to $7\times10^{-2}$ could be expected in this geometry, together with improved resolution. Such a large index change is comparable to that of phase-change materials~\cite{delaney-muskens2020family} and would substantially reduce the footprint or increase the achievable scale.

\subsection{Potential Non-Volatile Device}
\label{sec: Non-volatile}
A further extension of the programmable electro-optic device is to use photoconductive voltage control not only for index modulation but also for spatially programmable poling. In this concept, a DC external voltage is applied across the device stack, while an illumination pattern is projected onto the photoconductor. Unlike the devices above, here the DC field is antiparallel to the spontaneous polarization and exceeds the coercive field of LN in illuminated regions, reversing the local domain orientation, while dark regions remain below threshold. The illumination pattern therefore acts as a reconfigurable control for domain inversion.

The active core in such a device does not need to be lithium niobate. More generally, this concept can be extended to other ferroelectric or electro-optic $\chi^{(2)}$ materials, including lithium tantalate~\cite{yuEfficientTunableFrequency2025,sheltonRobustPolingFrequency2025}, barium titanate~\cite{aashnaPeriodicDomainInversion2024}, and related material systems. At present, this broader material implementation remains to be explored, and the poling conditions, material stability, and achievable spatial fidelity will depend on the specific ferroelectric response of the chosen core material.

The key distinction between this device and the devices described above is that the pattern inside the lithium niobate core is no longer temporary but fixed after poling, making it non-volatile. The reversed domains remain stable after the external voltage and illumination are removed. The illumination pattern therefore creates a non-volatile binary domain pattern within the device, with each region showing one of two opposite domain orientation states. Because the sign of the electro-optic response depends on the local polarization orientation, the two poled states would lead to opposite signs of the index change under an externally applied voltage. The device can therefore be regarded as a programmable binary core in which the local index change can be set to either positive or negative maximum and maintained without illumination.

In practice, the operating condition for such a programmable poling device requires the bright-state electric field in the active core to exceed the local coercive field, while the dark-state field must remain sufficiently below it. Therefore, we would need to optimize the stack to allow a sufficiently strong contrast and minimal field fringing. In that case the illumination pattern can create sharply defined inverted and non-inverted domains.

\subsection{Broader-Band Operation}
\label{sec:broadband}
In this work, we experimentally demonstrated an operating bandwidth of approximately \SI{100}{\nano\metre}, which is broader than that of most conventional inverse-designed photonic devices. This experimentally reported bandwidth is not set by an intrinsic limitation of the device itself, but is instead constrained by the tuning range of the tunable laser used in the measurement. 

In principle, the device can support a substantially larger operating bandwidth due to two key features of its architecture. First, the programmable refractive-index change provides relatively low index contrast compared with traditional inverse-designed photonic structures. Second, the transition between high- and low-index regions is intrinsically smooth, rather than abrupt, owing to the field fringing effect from the photoconductor voltage control and multilayer stack. These two characteristics reduce the strong wavelength sensitivity that typically arises in high-index-contrast nanophotonic structures with sharply defined interfaces, enabling much broader spectral operation.

\begin{figure}[!htbp]
    \centering
    \includegraphics[width=1\linewidth]{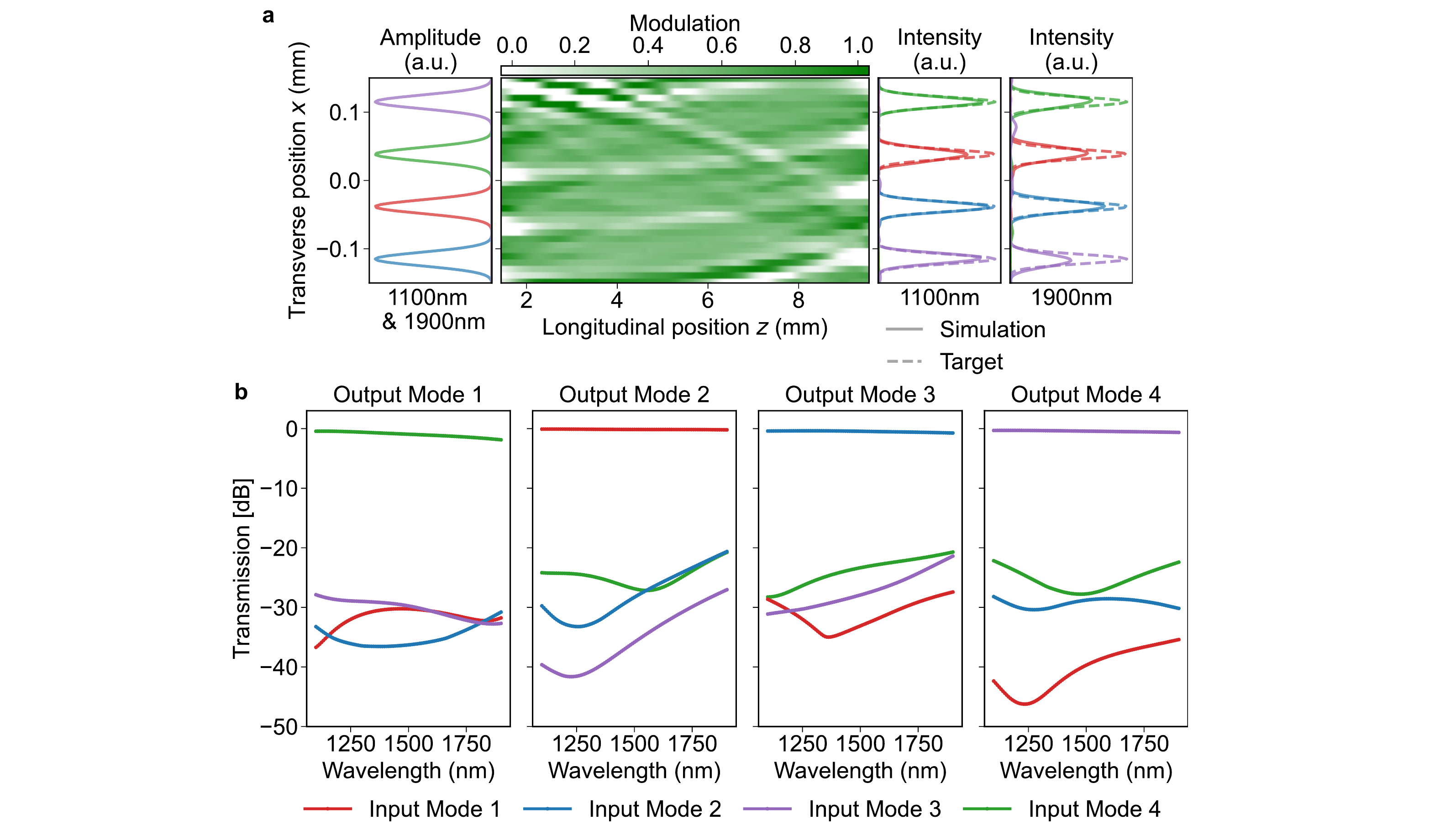}
    \caption{\textbf{Simulated $4\times4$ permutation task across the band from \SI{1100}{\nano\metre} to \SI{1900}{\nano\metre}.} \textbf{a}, Trained illumination pattern and corresponding output patterns at \SI{1100}{\nano\metre} and \SI{1900}{\nano\metre}. \textbf{b}, Transmission performance of the four \unitforN{}s over a wide bandwidth.}
    \label{fig:Broadband_simulation}
\end{figure}

Our simulations indicate that the operating bandwidth could extend to ${\sim}\SI{800}{\nano\metre}$, as shown in Fig.~\ref{fig:Broadband_simulation}. Such a bandwidth is large not only in comparison with conventional inverse-designed photonic devices, but also relative to existing programmable photonic devices. The present architecture therefore offers a route toward ultra-broadband programmable photonics, combining post-fabrication reconfigurability with a large spectral operating window.

\subsection{Potential Scale of Inverse-designed Functions with Improved Devices}
\label{sec:large scale}
The complexity of the transformation an inverse-designed device can realize is limited by the phase it can accumulate over its length. For a programmable weak index profile, this transformation is primarily controlled by the spatially varying refractive-index change, $\Delta n(x,z)$. For a fixed wavelength and comparable device geometry, the accumulated programmable phase scales as
\begin{equation}
    \Delta \phi \sim k_0 \int \Delta n(x,z)\, dz .
\end{equation}
Since our beam propagation equation (Eq.~\ref{eq:beam_propagation}) is invariant under the $x$, $z$ and $\Delta n(x,z)$ transformation,  for a similar optical function, the required propagation length $L_z$ scales inversely with the available programmable index contrast,
\begin{equation}
    L_z \propto \frac{1}{\Delta n}.
\end{equation}

We also note that the required transverse feature size $L_x$ would also be narrower, following a scaling of $L_x\propto 1/\sqrt{\Delta n}$. This simple scaling provides a useful estimate for how improved device parameters can either reduce the footprint of a fixed inverse-designed function or increase the dimensionality of the transformation achievable within a fixed footprint.

The largest function demonstrated here is an $8\times 8$ spatial permutation implemented on a $\sim\SI{1}{\centi\meter}\times\SI{1}{\centi\meter}$ programmable chip. The corresponding programmable refractive-index modulation is approximately $\Delta n = 1.7\times 10^{-3}$. If future device stacks increase the programmable index contrast to $\Delta n \sim 10^{-2}$, the available index modulation increases by a factor of $10^{-2}/(1.7\times 10^{-3}) \approx 6$. Therefore, a comparable optical transformation could in principle be realized in a device that is smaller by approximately the same linear factor. A function that now requires \SI{1}{\centi\meter} would then require ${\sim}\SI{1.7}{\milli\meter}$. This estimate is not intended as an exact design rule, because the final device size also depends on the number of supported modes, the desired crosstalk, the available transverse aperture, the smoothness of the index landscape, and the optimization objective. Nevertheless, it shows that increasing the programmable $\Delta n$ directly reduces the interaction length required to implement the same accumulated optical transformation.

Further improvements in programmable index modulation could lead to even more compact devices. For example, if future electro-optic or hybrid material stacks reach $\Delta n \sim 7\times 10^{-2}$, the modulation would be larger than the present value by a factor of $7\times 10^{-2}/(1.7\times 10^{-3}) \approx 41$, so under the same scaling argument, a function requiring a $\SI{1}{\centi\meter}$-scale interaction length in the present device could be shortened to a few hundred micrometers. This length scale is comparable to compact phase-change-material photonic devices~\cite{delaney-muskens2021nonvolatile,wu-li2024freeform,wuReconfigurableInversedesignedPhasechange2025}, while preserving the continuous reconfigurability and in situ trainability of the programmable electro-optic device. Other factors, such as the resolution of the projected illumination and of the input beam shaper, would also constrain the minimum size.

\begin{figure}[!htbp]
    \centering
    \includegraphics[width=1\linewidth]{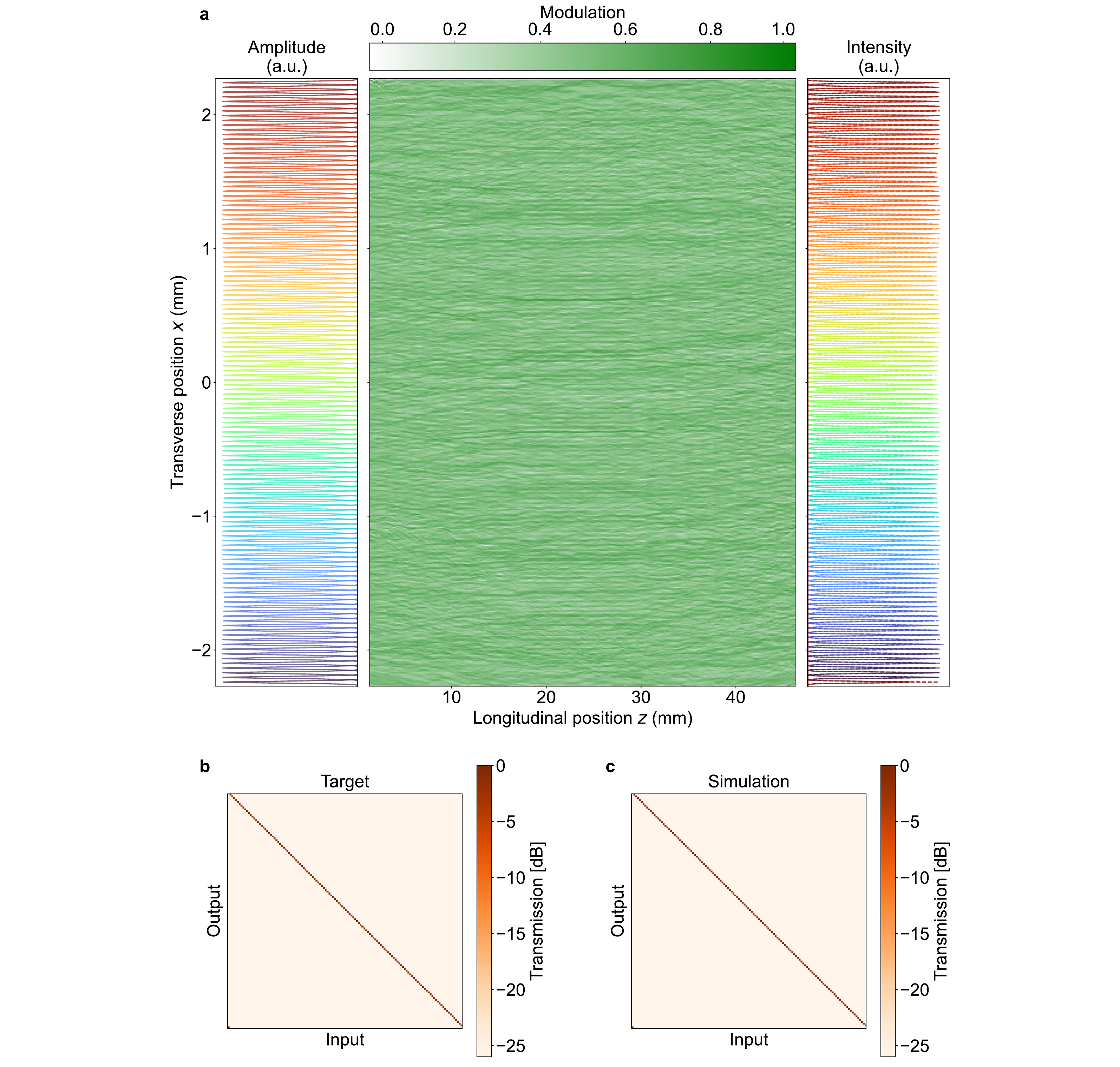}
    \caption{\textbf{Simulation results for $128\times128$ permutation task.} \textbf{a}, Trained illumination pattern and corresponding intensity outputs. Each color represents an input--output pair. The simulation assumed a programmable index change capability of $10^{-2}$, a device length of around \SI{5}{\centi\metre} and a working wavelength of \SI{1550}{\nano\metre}. \textbf{b}, Target transmission matrix of permutation task. \textbf{c}, Simulated transmission matrix of permutation task.}
    \label{fig:128Perm}
\end{figure}

\begin{figure}[!htbp]
    \centering
    \includegraphics[width=1\linewidth]{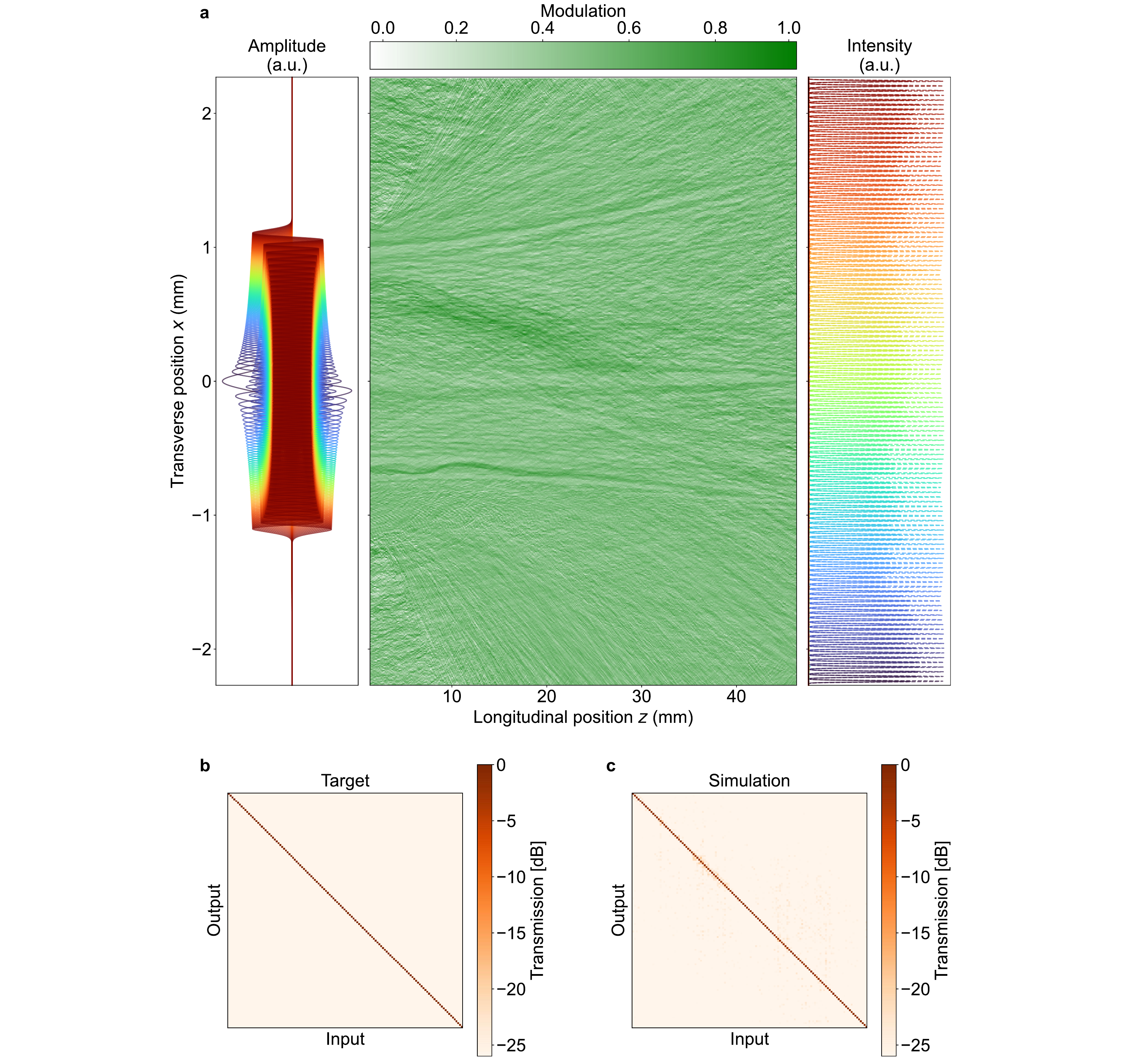}
    \caption{\textbf{Simulation results for $128\times128$ Hermite--Gaussian mode demultiplexer task.} \textbf{a}, Trained illumination pattern and corresponding intensity outputs. Each color represents an input--output pair. The simulation assumed a programmable index change capability of $10^{-2}$, a device length of around \SI{5}{\centi\metre} and a working wavelength of \SI{1550}{\nano\metre}. \textbf{b}, Target transmission matrix of mode demultiplexer task. \textbf{c}, Simulated transmission matrix of mode demultiplexer task.}
    \label{fig:128HGDemux}
\end{figure}

A second scaling scenario is to increase both the programmable index contrast and the device footprint. In the current experiments, a $\sim\SI{1}{\centi\meter}\times\SI{1}{\centi\meter}$ chip with $\Delta n \simeq 1.7\times 10^{-3}$ supports an experimentally trained $8\times 8$ permutation. In simulations of an improved device with $\Delta n \sim 10^{-2}$ and a footprint of approximately $\SI{5}{\centi\meter}\times\SI{5}{\centi\meter}$, the same programmable inverse-design approach supports a $128\times 128$ spatial mode permutation and a $128\times128$ Hermite--Gaussian mode demultiplexer. Details are shown in Fig.~\ref{fig:128Perm} and Fig.~\ref{fig:128HGDemux}. 

As a simple first-order extrapolation, we consider increasing the programmable index modulation from $\Delta n\sim10^{-2}$ in the simulated $128\times128$ device to $\Delta n\sim7\times10^{-2}$, corresponding to an additional factor of approximately seven in modulation strength. If all other device parameters are held fixed and the accessible transformation dimension is assumed to scale proportionally with the available index modulation, this extrapolation suggests that transformations involving on the order of $10^3$ input and output modes may become accessible within the same footprint. We have to note that this is a very rough estimate and it is only intended to show the potential scaling of the device, and it does not mean an experimentally validated $1000\times1000$ design. 

These estimates are intended to show the favorable scaling of a programmable inverse-designed photonic medium rather than to define a fundamental limit. Because the index profile is programmed after fabrication, scaling to larger and more complex transformations does not require more demanding lithography. Instead, complexity is transferred from nanofabrication to optical programming and in situ optimization. This difference is the key to the scalability of the device: the same physical device can implement in situ optimization to narrow the simulation--reality gap, potentially expanding the scale of the device beyond the current size limit of conventional inverse design.

\end{document}